\documentclass[fleqn,usenatbib]{mnras}

\usepackage{newtxtext,newtxmath}
\usepackage{graphicx}	
\usepackage{multicol} 
\usepackage{xcolor}
\usepackage{bm}	
\usepackage{pdflscape}
\usepackage{subcaption}
\usepackage[thinc]{esdiff}
\usepackage{amsmath}
\usepackage{soul}
\usepackage[T1]{fontenc}

\DeclareRobustCommand{\VAN}[3]{#2}
\let\VANthebibliography\thebibliography
\def\thebibliography{\DeclareRobustCommand{\VAN}[3]{##3}\VANthebibliography}

\graphicspath{{./}{figures/}}

\title[Magnetospheric Interactions in $\epsilon$ Lupi]{Discovery of Magnetospheric Interactions in the Doubly-Magnetic Hot Binary $\epsilon$ Lupi}

\author[A. Biswas et al.]{Ayan Biswas,$^{1,2,3}$\thanks{E-mail: ayan.biswas@queensu.ca}
Barnali Das,$^{4}$
Poonam Chandra,$^{5,1}$
Gregg A. Wade,$^{3}$
Matthew E. Shultz,$^{4}$ 
\newauthor Francesco Cavallaro,$^{6}$ \vspace{2mm}
Veronique Petit,$^{4}$ 
Patrick A. Woudt,$^{6}$
and Evelyne Alecian$^{7}$
\\ 
$^{1}$National Centre for Radio Astrophysics, Tata Institute of Fundamental Research, Ganeshkhind, Pune-411007, India\\
$^{2}$Department of Physics, Engineering Physics \& Astronomy, Queen’s University, Kingston, Ontario K7L 3N6, Canada\\
$^{3}$Department of Physics \& Space Science, Royal Military College of Canada, PO Box 17000 Station Forces, Kingston, ON K7K 0C6, Canada\\
$^{4}$Department of Physics and Astronomy, University of Delaware, Newark, DE 19716, USA\\
$^{5}$National Radio Astronomy Observatory, 520 Edgemont Road, Charlottesville VA 22903, USA\\
$^{6}$Inter-University Institute for Data Intensive Astronomy, Department of Astronomy, University of Cape Town, Private Bag X3, Rondebosch 7701, South Africa\\
$^{7}$Univ. Grenoble Alpes, CNRS, IPAG, 38000 Grenoble, France
}

\date{Accepted XXX. Received YYY; in original form ZZZ}
\pubyear{2022}

\begin{document}
\label{firstpage}
\pagerange{\pageref{firstpage}--\pageref{lastpage}}
\maketitle


\begin{abstract}
Magnetic fields are extremely rare in close, hot binaries, with only 1.5\% of such systems known to contain a magnetic star. The eccentric $\epsilon$ Lupi system stands out in this population as the only close binary in which both stars are known to be magnetic. We report the discovery of  strong, variable radio emission from $\epsilon$ Lupi using the upgraded Giant Metrewave Radio Telescope (uGMRT) and the MeerKAT radio telescope. The light curve exhibits striking, unique characteristics including sharp, high-amplitude pulses that repeat with the orbital period, with the brightest enhancement occurring near periastron. The characteristics of the light curve point to variable levels of magnetic reconnection throughout the orbital cycle, making $\epsilon$ Lupi the first known high-mass, main sequence binary embedded in an interacting magnetosphere. We also present a previously unreported enhancement in the X-ray light curve obtained from archival XMM-Newton data. The stability of the components' fossil magnetic fields, the firm characterization of their relatively simple configurations, and the short orbital period of the system make $\epsilon$ Lupi an ideal target to study the physics of magnetospheric interactions. This system may thus help us to illuminate the exotic plasma physics of other magnetically interacting systems such as moon-planet, planet-star, and star-star systems including T Tauri binaries, RS CVn systems, and neutron star binaries.
\end{abstract}



\section{Introduction}

A small fraction of hot, massive stars have been found to harbor extremely stable, ordered (usually dipolar) magnetic fields, which are of $\sim$ kG strength \citep{Wade2014, Grunhut2017}. The presence of such organized surface magnetic fields can channel and confine the outflowing stellar winds \citep{Babel1997}, creating a magnetosphere that can radiate in various wavebands \citep{Andre1988, Linsky1992}. The confinement of the wind in the magnetosphere leads to the suppression of mass loss \citep{ud-Doula2002, Petit2017}. The magnetic field thus plays a crucial role in deciding the mass of the magnetic star at the end of the main sequence \citep{Keszthelyi2019, Keszthelyi2020}.
\cite{Petit2017} have shown that magnetospheric mass-loss suppression is a possible formation channel for the production of the heavy stellar-mass black holes such as those detected by gravitational wave observatories \citep{Abbott2016a, Abbott2016b}.

\begin{figure*}
    \centering
    \begin{multicols}{2}
        \subcaptionbox{\label{fig:1x}}{\includegraphics[width=0.88\linewidth]{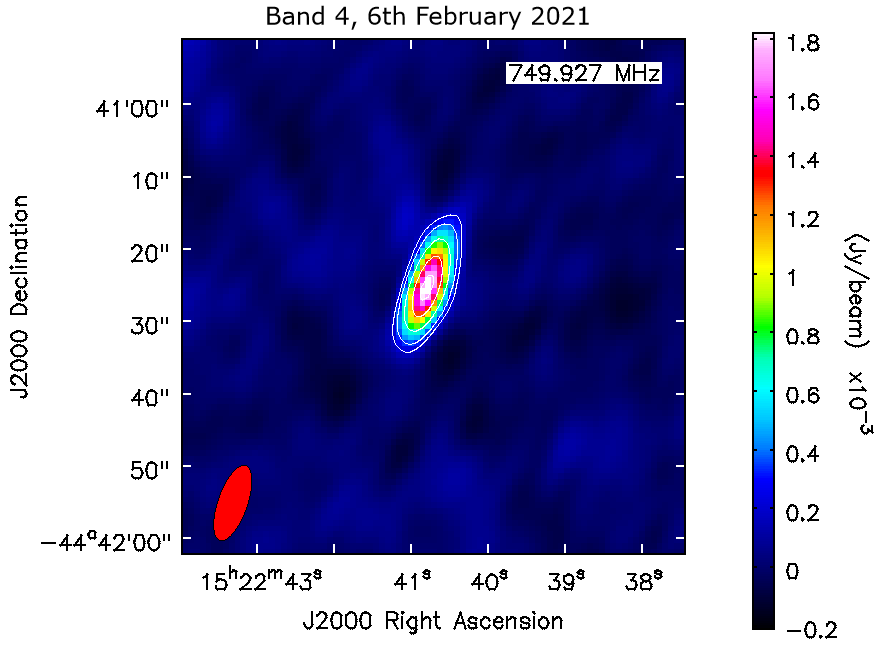}}\par 
        \subcaptionbox{\label{fig:2x}}{\includegraphics[width=0.9\linewidth]{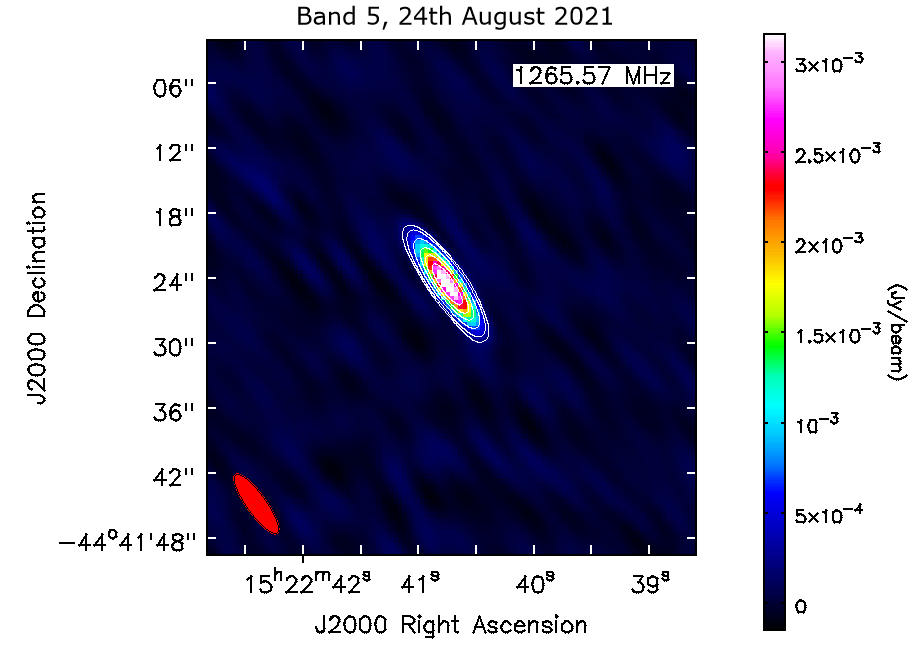}}\par 
    \end{multicols}
\caption{uGMRT continuum images of $\epsilon$ Lupi. Panel (a) shows the band 4 (550--900 MHz) detection on 6 February 2021 (orbital phase $\phi_{\rm orb} = 0.165$) and panel (b) shows a band 5 (1050--1450 MHz) detection on 24 August 2021 ($\phi_{\rm orb}$ = 0.918). The synthesized beam is shown in red filled ellipses (size $8.37\arcsec \times 3.09\arcsec$ for band 4 and $7.05\arcsec \times 1.99\arcsec$ for band 5) at the bottom left of each plot. The contours in white show the detection signiﬁcance and are drawn at 6, 10, 20, 35, 50 $\sigma$ for band 4, and 6, 10, 20, 35, 50, 70 $\sigma$ for band 5, where $\sigma$ is the rms noise of the corresponding images near the phase center ($\sigma \sim \ 36 \ \mu \rm Jy$/beam for band 5 and $\sim 35 \ \mu \rm Jy$/beam for band 4).\label{fig:radio_image}}
\end{figure*}

Binarity impacts stellar populations in numerous ways, such as by modifying surface abundances, enriching the interstellar medium, and affecting the demise of massive stars as supernova and $\gamma$-ray burst explosions \citep{Izzard2006}. In the  Binarity and Magnetic Interactions in various classes of Stars (BinaMIcS)    survey, about 170 short-period, double-lined spectroscopic intermediate-mass and high-mass binaries with orbital periods of less than 20 days were observed with high-resolution spectropolarimeters, of which only 1.5\% were found to contain a magnetic star
\citep{Alecian2015}. This fraction is much lower than the percentage of isolated intermediate-mass and high-mass magnetic stars ($\sim 10\%$, \citealt{Grunhut2017, Sikora2019}). Among the magnetic binaries of this survey, magnetic fields have been detected in only one of the two components, with the sole exception of $\epsilon$ Lupi A (henceforth referred to as $\epsilon$ Lupi). \cite{Shultz2015} reported dipolar surface magnetic field strengths ($B_{\rm d}$) of roughly 0.8 kG and 0.5 kG in the primary and secondary components of $\epsilon$ Lupi, respectively.

Apart from $\epsilon$ Lupi, there exist just two other reported doubly-magnetic hot binary systems, HD 156424 \citep{Shultz2021} and BD +40 175 \citep{Elkin1999,Semenko2011}. Both systems have long (years to decades) orbital periods, implying that their stellar components evolve practically as single stars. In contrast, the short orbital period  ($P_{\rm orb} \sim$ 4.6 days) of $\epsilon$ Lupi implies the possibility of significant tidal interactions  and potentially magnetospheric interactions as well \citep{Pablo2019}.

In recent years, extensive studies have been performed of magnetism in intermediate-mass and high-mass stars, aiming at characterizing various magnetospheric emissions such as H$\alpha$, X-ray, UV, IR, and non-thermal radio emission \citep{ Shultz2020, Naze2014, Erba2021, Oksala2015, Leto2021, Das2022b, Das2022}. These studies together have been able to provide a plethora of information regarding the three-dimensional structure of their magnetospheres \citep{Townsend2005, ud-Doula2013, Leto2016, Das2021}. The effect of binarity in the evolution of massive stars has been studied in great detail \citep[e.g.][]{Sana2012}. But the study of the combined effects of binarity and magnetic fields on massive star magnetospheres is in its infancy (e.g. \citealt{Shultz2018}). The doubly magnetic massive binary system $\epsilon$ Lupi  provides us with a unique test-bed to examine such effects. Studying this system may help understand similar magnetically interacting binaries e.g. moon-planet, planet-star, and star-star systems including T Tauri binaries \citep{Tofflemire2017, Salter2010, Massi2006}, RS CVn systems \citep{Karmakar2023, Uchida1983}, and neutron star binaries \citep{Most2022a, Most2022b, Cherkis2021, Wang2018}.

\begin{table}
\caption[Stellar and orbital parameters of epsilon lupi]{Different stellar and orbital parameters of $\epsilon$ Lupi. Parameters with subscript `$\rm phot$' represent parameters obtained from photometric observations by \cite{Pablo2019}, whereas parameters with subscript `$\rm spec$' represent parameters obtained from spectroscopic observations by \cite{Shultz2019}.}  References: $a=$ \cite{Pablo2019}; $b=$ \cite{Shultz2019};  $c=$ \cite{Shultz2015};  $d=$ \cite{Shultz2018}. \label{tab:parameters}
\begin{tabular}{lcc}
\hline
\hline
{Parameters}   & {Primary} & {Secondary} \\ \hline
 \multicolumn{3}{|c|}{\textit{Stellar Parameters}} \\
 Mass ($M_{\rm phot}/M_{\odot}$)$^a$  &  $11.0_{-2.2}^{2.9}$ & $9.2_{-1.9}^{2.4}$ \\
 Mass ($M_{\rm spec}/M_{\odot}$)$^b$ & 7.7  $\pm$0.5 & 6.4 $\pm$0.3 \\
 Radius ($R_{\rm phot}/R_{\odot}$)$^a$ &  $4.64_{-0.48}^{0.37}$ &  $4.83_{-0.46}^{0.42}$ \\
 Radius ($R_{\rm spec}/R_{\odot}$)$^b$ &   5.1 $\pm$0.07 & 3.7 $\pm$0.5  \\
Luminosity ($L/L_{\odot}$)$^a$  & $3407_{-567}^{658}$ & $2197_{-399}^{489}$ \\
Effective Temperature, $T_{\rm eff}$  (K)$^a$  & 20500 & 18000 \\
$v \sin i$ ($\rm km s^{-1}$)$^d$ & 35 $\pm$ 10 & 35 $\pm$ 5 \\
Obliquity, $\beta$ ($^{\circ}$)$^a$ & $< 30.7$ & $< 21.4$ \\
Dipolar Field Strength, $B_{\rm P}$ (kG)$^a$ & $>0.79$ & $>0.5$ \\
$\log (R_{\rm A}/R_{\rm K})$$^a$  & 0.63 $\pm$0.07 & 0.22 $\pm$0.08 \\
 \multicolumn{3}{|c|}{\textit{Orbital Parameters}} \\
Orbital Period, $P_{\rm orb}$ (days)$^a$ &   \multicolumn{2}{|c|}{$4.559646^{0.000005}_{-0.000008}$} \\
$T_0$ (HJD - 2400000)$^a$ &   \multicolumn{2}{|c|}{$2439379.875^{0.024}_{-0.019}$} \\
Eccentricity (e)$^a$ &   \multicolumn{2}{|c|}{$0.2806^{0.0059}_{-0.0047}$} \\
Semi-major axis, a  ($R_{\odot}$)$^a$ &   \multicolumn{2}{|c|}{$31.5^{2.5}_{-2.3}$} \\
Apsidal motion $\diff{\omega}{t}$ ($^{\circ}$/yr)$^a$ &   \multicolumn{2}{|c|}{$1.1^{0.1}_{-0.1}$} \\ \hline
\end{tabular}
\end{table}

\subsection[Epsilon Lupi]{$\epsilon$ Lupi}

\begin{table*}
 \caption{Observation details and imaging results of uGMRT observations. For each observing day, only mean phases are mentioned for simplicity. The bold faced phases represent periastron observations (defined by phase $0 \pm 0.02$). As the observations during May 2022 are of short-duration, flux density values for small-duration scans are not shown individually in those cases.  \label{tab:flux}}
 \begin{tabular}{ccccccc}
  \hline
\hline
{Obs. Date} & {Mean Phase} &  {Heliocentric Julian Date} & \multicolumn{2}{c}{Band 5} & \multicolumn{2}{c}{ Band 4 }  \\
{} & {$\phi_{\rm orb}$} &  {(HJD-2459000)} &
{F$_{1250}$ (mJy)} & { rms ($\mu$Jy/beam)} & {F$_{750}$ (mJy)} & { rms ($\mu$Jy/beam)} 
 \\
  \hline
02 Jan 2021  & 0.340 $\pm$ 0.003 & 216.64407 $\pm$ 0.01320  & 2.905 $\pm$ 0.045 & 16.76 & 1.828 $\pm$ 0.084 & 37.87 \\
  & 0.349 $\pm$ 0.003 &  216.68194 $\pm$ 0.01320  & 2.754 $\pm$ 0.049  & 17.86 &  1.882 $\pm$ 0.072 & 38.18    \\
  & 0.355 $\pm$ 0.002 &  216.70935 $\pm$ 0.00727  & 2.872 $\pm$ 0.082  & 21.09 &  1.844 $\pm$ 0.093 & 44.61    \\
  \hline
05 Jan 2021 &   \textbf{0.000 $\pm$ 0.003} & 219.64982 $\pm$ 0.01320  & 4.270 $\pm$ 0.110 & 38.99 & 2.003 $\pm$ 0.069 & 38.72 \\
  & \textbf{0.007 $\pm$ 0.003} & 219.68246 $\pm$ 0.01320  & 4.520 $\pm$ 0.130  & 40.98 &  2.069 $\pm$ 0.076 & 35.63    \\
  & \textbf{0.014 $\pm$ 0.003} & 219.71649 $\pm$ 0.01320  & 4.600 $\pm$ 0.100  & 36.60 &  2.058 $\pm$ 0.082 & 40.13    \\
  & \textbf{0.020 $\pm$ 0.001} & 219.74288 $\pm$ 0.00556  & 4.440 $\pm$ 0.110  & 49.80 &  2.047 $\pm$ 0.091 & 45.20    \\
    \hline
08 Jan 2021  & 0.651 $\pm$ 0.003 & 222.61876 $\pm$ 0.01320  & 2.797 $\pm$ 0.045 & 21.02 & 1.630 $\pm$ 0.029 & 29.21 \\
  & 0.658 $\pm$ 0.003 & 222.65279 $\pm$ 0.01320  & 2.770 $\pm$ 0.077  & 36.14 &  ... & ...    \\
  & 0.666 $\pm$ 0.003 & 222.68613 $\pm$ 0.01320  & 2.840 $\pm$ 0.056  & 24.61 &  ... & ...    \\
  & 0.674$\pm$ 0.004  & 222.72363 $\pm$ 0.01667  & 2.500 $\pm$ 0.063  & 27.18 &  ... & ...    \\
    \hline
15 Jan 2021  & 0.186 $\pm$ 0.003 & 229.61876 $\pm$ 0.01320  & 2.230 $\pm$ 0.039 & 22.51 & 1.268 $\pm$ 0.072 & 35.97 \\
  & 0.193 $\pm$ 0.003 & 229.65279 $\pm$ 0.01320  & 2.289 $\pm$ 0.038  & 21.69 &  1.272 $\pm$ 0.066 & 34.07    \\
  & 0.201 $\pm$ 0.003 & 229.68613 $\pm$ 0.01320  & 2.243 $\pm$ 0.040  & 24.31 &  1.302 $\pm$ 0.074 & 35.15    \\
  & 0.209 $\pm$ 0.001 & 229.72363 $\pm$ 0.00556  & 2.185 $\pm$ 0.045  & 27.93 &  1.236 $\pm$ 0.077 & 45.42    \\
    \hline
18 Jan 2021  & 0.836 $\pm$ 0.003 & 232.58476 $\pm$ 0.01320  & 2.140 $\pm$ 0.053 & 22.51 & 1.547 $\pm$ 0.054 & 32.50 \\
  & 0.844 $\pm$ 0.003 & 232.62014 $\pm$ 0.01320  & 2.125 $\pm$ 0.035  & 24.44 &  1.629 $\pm$ 0.042 & 35.08    \\
  & 0.851 $\pm$ 0.003 & 232.65347 $\pm$ 0.01320  & 2.093 $\pm$ 0.041  & 25.81 &  1.426 $\pm$ 0.070 & 51.29    \\
  & 0.858 $\pm$ 0.002 & 232.68264 $\pm$ 0.00903  & 1.942 $\pm$ 0.040  & 25.39 &  1.401 $\pm$ 0.067 & 54.17    \\
    \hline
06 Feb 2021  & \textbf{0.010 $\pm$ 0.006} & 251.49313 $\pm$ 0.02573  & ... & ... & 1.978 $\pm$ 0.073 & 34.69 \\
  \hline
22 Feb 2021  & 0.491 $\pm$ 0.003 & 267.48709 $\pm$ 0.01250  & 2.299 $\pm$ 0.045 & 24.26 & 1.414 $\pm$ 0.120 & 63.02 \\
  & 0.498 $\pm$ 0.003 & 267.51973 $\pm$ 0.01250  & 2.235 $\pm$ 0.059  & 24.53 &  1.375 $\pm$ 0.063 & 37.03    \\
  & 0.506 $\pm$ 0.003 & 267.55515 $\pm$ 0.01250  & 2.088 $\pm$ 0.046  & 25.38 &  1.279 $\pm$ 0.072 & 37.10    \\
  & 0.513 $\pm$ 0.003 & 267.58988 $\pm$ 0.01459  & 2.155 $\pm$ 0.042  & 26.04 &  1.168 $\pm$ 0.063 & 37.19    \\
    \hline
20 Jul 2021  & 0.070 $\pm$ 0.002 & 416.04451 $\pm$ 0.01041  & 2.899 $\pm$ 0.092 & 40.67 & 1.380 $\pm$ 0.100 & 57.46 \\
  & 0.076 $\pm$ 0.002 & 416.07228 $\pm$ 0.01041  & 2.720 $\pm$ 0.110  & 38.43 &  1.630 $\pm$ 0.170 & 91.58    \\
  & 0.082 $\pm$ 0.002 & 416.09589 $\pm$ 0.00694  & 2.624 $\pm$ 0.099  & 41.71 &  1.310 $\pm$ 0.110 & 63.83    \\
    \hline
24 Aug 2021 &   0.743 $\pm$ 0.002 & 451.02847 $\pm$ 0.01041  & 3.459 $\pm$ 0.058 & 29.85 & 1.790 $\pm$ 0.160 & 65.87 \\
  & 0.750 $\pm$ 0.002 & 451.06308 $\pm$ 0.01041  & 3.538 $\pm$ 0.062  & 30.75 &  1.460 $\pm$ 0.140 & 70.28    \\
  & 0.7564 $\pm$ 0.002 & 451.09155 $\pm$ 0.01041  & 3.496 $\pm$ 0.069  & 44.32 &  0.967 $\pm$ 0.128 & 80.25    \\
    \hline
08 Sep 2021  & \textbf{0.010 $\pm$ 0.007} & 465.96733 $\pm$ 0.03330  & 4.490 $\pm$ 0.023 & 24.60 & ... & ... \\ \hline
20 May 2022  & 0.785 $\pm$ 0.005 & 720.25251 $\pm$ 0.02361 & 2.439 $\pm$ 0.052 & 30.71 & 1.473 $\pm$ 0.088 & 61.90 \\  \hline 
26 May 2022 & 0.097 $\pm$ 0.008 & 726.22645 $\pm$ 0.03710 & 4.102 $\pm$ 0.068 & 31.20 & ... & ... \\ \hline
29 May 2022 & 0.749 $\pm$ 0.005 & 729.20490 $\pm$  0.02396 & 3.685 $\pm$ 0.083 & 49.95 & 1.660 $\pm$ 0.160 & 63.61 \\ \hline
30 May 2022 & 0.965 $\pm$ 0.005  & 730.19134 $\pm$ 0.02361 & 3.359 $\pm$ 0.085 & 37.36 & ... & ... \\ 
\hline
\end{tabular}
\end{table*}

$\epsilon$ Lupi is a double-lined spectroscopic binary (SB2) in a close orbit \citep{Uytterhoeven2005} where the component spectral types are identified as B3IV and B3V \citep{moore1911, campbell1928, Thackeray1970}. \cite{Pablo2019} discovered a `heart-beat' type light curve from photometric observations obtained with the BRIght-star Target Explorer (BRITE) constellation of space telescopes \citep{Weiss2014}, which they interpreted as a result of tidal distortion. Such light curves are normally observed in systems with high eccentricities, and can be used to determine different orbital parameters, and to directly measure masses and radii. \cite{Pablo2019} determined an eccentric orbit ($e=0.28$) with an orbital period of approximately 4.6 days. The primary and the secondary possess comparable masses ($M_{\rm P}=11.0 \ M_{\odot}$ \& $M_{\rm S}= 9.2 \ M_{\odot}$), radii ($R_{\rm P}=4.6 \ R_{\odot}$ \& $R_{\rm S}= 4.8 \ R_{\odot}$) and effective temperatures ($T_{\rm eff,P}= 20.5$ kK \& $T_{\rm eff,S}= 18$ kK) (Table \ref{tab:parameters}). The orbit of $\epsilon$ Lupi is characterized very well with the help of numerous radial velocity measurements \citep{Thackeray1970, Uytterhoeven2005, Pablo2019}.  However, the rotational periods of the two components are not known. The longitudinal magnetic fields $\langle B_z \rangle$ of the components show weak temporal modulation that points toward the magnetic axes being approximately aligned with the rotational axes (Fig. \ref{fig:lc}, bottom). The constant positive and negative $\langle B_z \rangle$ for the secondary and primary, respectively, shows that the field axes are anti-aligned. { In such a scenario, where the rotation axes are aligned, magnetic axes are anti-aligned, and the obliquity of the field with respect to the rotation axis in each star is assumed to be small, the energy due to the magnetic dipole–dipole interaction force is at a minimum, making this a stable configuration \citep{Pablo2019}. }

To obtain the radio light curve with respect to orbital phase we used the following ephemeris reported by \cite{Pablo2019}:

\begin{equation}
    {\rm HJD}  = T_0  + P_{\rm orb}E \ ,
\end{equation}

\noindent
where $E$ is the phase of the observation, $P_{\rm orb}=4.559646^{0.000005}_{-0.000008}$ is the orbital period, and $T_0 = 2439379.875^{0.024}_{-0.019}$ is defined by the reference periastron phase. The system has a significant apsidal motion with ${\rm d}\omega/{\rm d}t = 1.1 \pm 0.1^{\circ}$/yr \citep{Pablo2019}. Orbital phases calculated in this article account for apsidal motion unless stated otherwise.

\cite{Shultz2015} predicted that the magnetospheres of the two stars are always overlapping, and that the strength of the interaction should change over the orbital cycle due to the eccentricity (Table \ref{tab:parameters}). The system was undetected in H$\alpha$ and UV \citep{Petit2013, Shultz2019}; but it was detected in X-rays with XMM-Newton (\cite{Naze2014}, see section \ref{xmm_obs} for details) and NICER \citep{Das2023}. In this paper, we first time report the radio observations and detection of $\epsilon$ Lupi system.
Since radio emission is common in magnetic hot stars \citep{Leto2021, Shultz2022}, radio observations 
may be the best way to probe its magnetosphere. The magnetospheric interactions that one would expect due to overlapping magnetospheres may be witnessed as enhancements in radio and/or X-ray light curves. In this paper, we report the discovery of radio emission and present a detailed radio study of $\epsilon$ Lupi with the upgraded Giant Metrewave Radio Telescope \citep[uGMRT,][]{Swarup1992, Gupta2017} and the MeerKAT radio telescope \citep{Jonas2009}. We also perform the first time series analysis of the archival XMM-Newton data of 8 ksec observation.

This paper is structured as follows: The observation details and data analysis procedures are explained in section \ref{sec:Observations}. In section \ref{sec:Results} we present results from the radio observations. In section \ref{sec:Discussion} we explore a variety of models in order to interpret the results in the context of magnetospheric interaction. Finally, in section \ref{sec:Conclusion} we summarize our conclusions.

\section{Observations and Data Reduction} \label{sec:Observations}

\subsection{The uGMRT}

As binarity may lead to variations in the radio flux density on the timescale of the orbital period, we sampled various phases across the orbital cycle of $\epsilon$ Lupi ($P_{\rm orb} \sim$ 4.56 days) with the uGMRT to check for variability in the radio emission. In addition, to obtain spectra at different phases, we performed simultaneous measurements in band 5 (1050--1450 MHz) and band 4 (550--900 MHz) in a sub-array mode where each array contained roughly 14 antennae distributed along the Y-shaped arms of uGMRT. The sub-array mode was chosen with an intention to avoid the possibility of spurious results regarding spectral properties over band 4 and band 5 due to time-variability (if there is any). Initially, the target was observed at six different orbital phases.  The observations were taken between 2 January 2021 and 22 February 2021 (see Table \ref{tab:flux}) during uGMRT Cycle 39   (ObsID: 39\textunderscore 086, PI: A. Biswas). Three additional observations were taken between 20 July 2021 and 8 September 2021 under Director's Discretionary Time (DDT) proposal (ObsID: ddtC\textunderscore 189). Finally, four additional follow-up observations are reported here that were taken during { May 2022, under} uGMRT Cycle 42  (ObsID: 42\textunderscore 018, PI: A. Biswas). The 6 February and 8 September (2021) observations were not carried out in sub-array mode, and all 30 antennae of uGMRT were used in the band 4 and the band 5 frequency settings, respectively.

Each observation was of approximately 4 hours, with an on-source time of  100 -- 150 min. For all observations, visibilities were recorded in full polar mode with a bandwidth of 400 MHz divided into 2048 frequency channels in both frequency bands. The standard flux calibrator J1331+305  (also known as 3C286) was observed at the beginning (and sometimes also towards the end) for 10--15 minutes to calibrate the absolute flux density scale. Each scan of the target was preceded and followed by a 5-minute observation of the phase calibrator J1626-298 chosen from the Very Large Array calibrator list such that it is located within 15$^{\circ}$ of $\epsilon$ Lupi. The phase calibrator was observed to track instrumental phase and gain drifts, atmospheric and ionospheric gain, and also for monitoring the data quality for spotting occasional gain and phase jumps due to instrumental anomalies.

\begin{figure}
    \centering
        \subcaptionbox{\label{fig:2a}}{\includegraphics[width=0.87\linewidth]{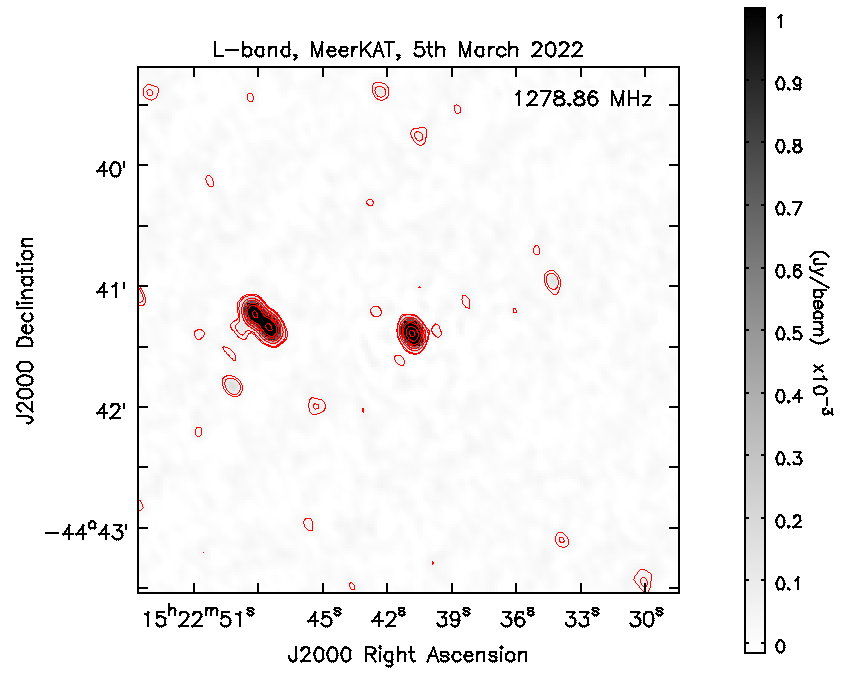}}\par 
        \subcaptionbox{\label{fig:2b}}{\includegraphics[width=0.91\linewidth]{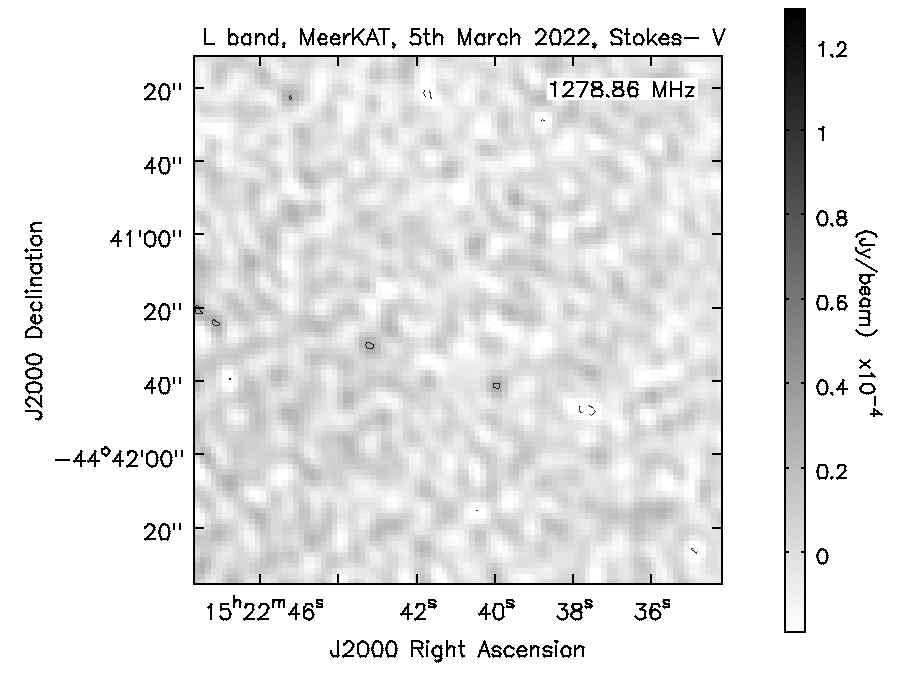}}\par 
        \subcaptionbox{\label{fig:2c}}{\includegraphics[width=0.91\linewidth]{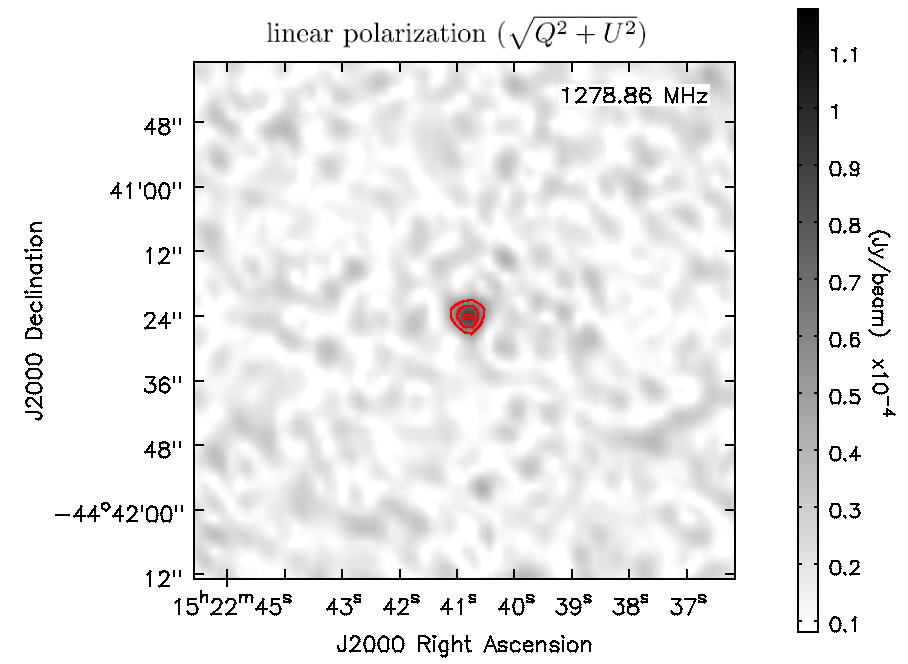}}\par 
\caption{(a) MeerKAT L-band Stokes I image of the periastron observation on 5 March 2022. The contours in red show the detection signiﬁcance and are drawn at 3, 5, 15, 30, 60, 90, 180, 300 $\sigma$, where $\sigma$ is the rms noise of the Stokes I image near the phase center (here, 16 $\mu$Jy). (b) Stokes V image of the periastron phase showing non-detection. The rms noise in the shown Stokes V image is $\sigma \sim$ 9 $\mu$Jy, thus constraining the upper limit of the Stokes V flux of our source to be 27 $\mu$Jy. Contours are drawn at 3 $\sigma$.  (c) Linear polarization map { obtained from Stokes Q and U images} confirming the presence of linear polarization. The contours in magenta are drawn at 3, 4, 5 $\sigma$, where $\sigma$= 17 $\mu$Jy/beam. The fitted peak flux from the full band is 90 $\mu$Jy, showing a linear polarization fraction of 1.6\%.\label{fig:meerkat_image}}
\end{figure}

\begin{table*}
 \caption{Observation details and imaging results of the MeerKAT observations. { The bold faced phases represent periastron observations (defined by phase $0 \pm 0.02$). The flux-densities are not offset-corrected in this table (for all Stokes parameters). \iffalse The flux densities for the full band obtained from our analysis are very similar to the flux densities obtained from the pipeline data products available in MeerKAT archive.  \fi }  \label{tab:meerkat_flux}}
 \begin{tabular}{cccccccc}
  \hline
\hline
{Obs. Date} &  {$\phi_{\rm orb}$} & {HJD-2459000} &   { Band (MHz)} & {Stokes} &  {Flux Den. (mJy)} &   {rms ($\mu$Jy/beam)}  & {Pol. Fraction} \\ \hline
13 Feb 2022 &  0.609 & 623.6915 $\pm$ 0.0306  & 1032-1455 & I  & 5.581 $\pm$ 0.045 & 20.84 & ... \\
 &   &   & 856-1717 & I   & 5.981 $\pm$ 0.034 & 25.04 & ...  \\ \hline
04 Mar 2022 & 0.759  & 642.6137 $\pm$ 0.0945  & 856-1717 & I & 5.393 $\pm$ 0.071 & 14.19 & ... \\ \hline
05 Mar 2022 & \textbf{0.00}  & 643.6992 $\pm$ 0.0945  & 856-1717 & I  & 5.746 $\pm$ 0.010 & 14.85 & ... \\
 & & & & V & $<$ 0.028 &  9.4 & $<$ 0.49 \% \\
  & & & & Q & $<$ 0.035 & 11.58 & $<$ 0.61 \% \\
 & & & & U & 0.094 $\pm$ 0.005 & 9.89 & {  1.64 $\pm$ 0.09 \% } \\
 &   &   & 1060-1450 & I  & 5.769 $\pm$ 0.043 & 23.91 & ...  \\
  & & & & V & $<$ 0.045 &  15.1 & $<$ 0.78 \%  \\
  & & & & Q & $<$ 0.041 &  13.82 & $<$ 0.71 \%  \\
  & & & & U & 0.118 $\pm$ 0.030 &  20.38 & { 2.05 $\pm$ 0.52 \% } \\ \hline
12 Mar 2022 &  0.526 & 650.6748 $\pm$ 0.0306  & 856-1717 & I  & 3.229 $\pm$ 0.038 & 23.42 & ... \\
 &   &   & 1032-1455 & I   & 3.167 $\pm$ 0.043 & 22.40 & ...  \\ \hline

\end{tabular}
\end{table*}

The data were analyzed using the Common Astronomy Software Applications (CASA) package \citep{McMullin2007}.  Bad channels and dead antennae were identified and flagged using the tasks `{\it plotms}' and `{\it flagdata}'. Radio Frequency Interference (RFI) was removed by running the automatic flagging algorithm “{\it tfcrop}” on the uncalibrated dataset. The edges of the bands were flagged manually due to their low gains. We performed bandpass calibration (task `{\it bandpass'}) using the flux calibrator 3C286 to obtain  frequency-dependent antenna gains, whereas to get the time-dependent gains we used the task `{\it gaincal}'. These calibrations (along with delay and flux calibration) were applied to all the sources and the corrected data were inspected. The automatic RFI excluding algorithm `{\it rflag}' was used on the calibrated dataset, and further flagging was done manually using the task `{\it flagdata}'. This  flagging $+$ calibration step was repeated cyclically until the corrected data did not show significant RFI.

The calibrated data for the target $\epsilon$ Lupi were then averaged over 4 frequency channels resulting in a final spectral resolution of 0.78 MHz. The typical final bandwidth for our observations was 290 MHz for band 4 and 370 MHz for band 5. The imaging was done using the CASA task `{\it tclean}' with deconvolver `{\it mtmfs}' (Multiscale Multi-frequency with W-projection, \citealt{Rau2011}). To improve the imaging quality, several rounds of phase-only and two rounds of amplitude and phase (A \& P)-type self-calibration were done using the `{\it gaincal}' \& `{\it tclean}' tasks. Sample images of band 4 and band 5 are shown in Fig. \ref{fig:radio_image}. In order to inspect time-variability, these self-calibrated data-sets were divided into smaller time-slices and the imaging was done for 5-15 minute time-slices. Table \ref{tab:flux} reports the flux densities from images obtained by dividing each observation into several scans. 

\subsection{MeerKAT}

We observed $\epsilon$ Lupi with MeerKAT between 13 February and 12 March 2022 under the DDT project DDT-20220120-AB-01 (PI: A. Biswas). MeerKAT is equipped with dual (linearly) polarized L-band (900 - 1670 MHz) and UHF band (580 -  1015 MHz) receivers in all the antennae. This allowed us to  obtain the polarization information as well as to cover a broader wavelength range. One set of observations was performed during periastron phase spanning 5 hours (Table \ref{tab:meerkat_flux}), one set near the 0.75 phase (as another follow-up observation) spanning 5 hours, and finally two 2-hour observations were taken at two random phases { (phases 0.609 and 0.526)}. During each observation, J1939-6342 was used as the flux calibrator,  J1130-1449 was used as the polarization calibrator, and J1501-3918 was used as the gain calibrator. Similar to the uGMRT observations, sub-array mode was  utilized during these observations with 32 antennae in each band. The data were recorded in full polarization mode with a bandwidth of 856 MHz divided into 4096 channels. In this paper, a detailed analysis from only the L-band is discussed. { The UHF band analysis will be presented in a later publication.}

The data were analyzed using the {\it processMeerKAT} script \citep{Collier} available on the {\it ilifu} \footnote{\url{https://www.ilifu.ac.za/}} cluster. To produce a reliable wideband continuum and polarization calibration, and to decrease the run-time, the script splits the band into several spectral windows (SPWs) and solves for each SPW separately. Several rounds of self-calibration steps were performed, and the final image was made using the `{\it tclean}' task in CASA. To compare the MeerKAT results with the uGMRT, the calibrated datasets were split to match the frequency range of band 5 of uGMRT. Images corresponding to different Stokes parameters (Stokes I, Q, U, and V) were then obtained. For the periastron phase, the data were imaged in smaller time ranges or single SPWs to obtain the light curve and spectra, respectively.

We noticed a nearly constant offset between the flux densities of sources in the field of view obtained from MeerKAT and uGMRT at all epochs. We selected some of the sources from the field of view for which the flux density does not vary significantly within different days obtained from the same telescope. We then obtained the mean offset value (Table \ref{tab:flux_mismatch}) and corrected the MeerKAT flux density accordingly. 

\subsection{XMM-Newton} \label{xmm_obs}

We used the archival  XMM-Newton X-ray data for $\epsilon$ Lupi taken on 2013 March 4 (ObsID: 0690210201, PI: Y. Nazé). During the 7919 s observation, both EPIC and RGS instruments were turned on. The thick filter was used while operating in the full-frame mode of EPIC instruments and default spectroscopy mode of RGS instruments. The optical monitor (OM) was turned off as the source is optically bright.

\begin{figure}%
\includegraphics[{width=0.455\textwidth}]{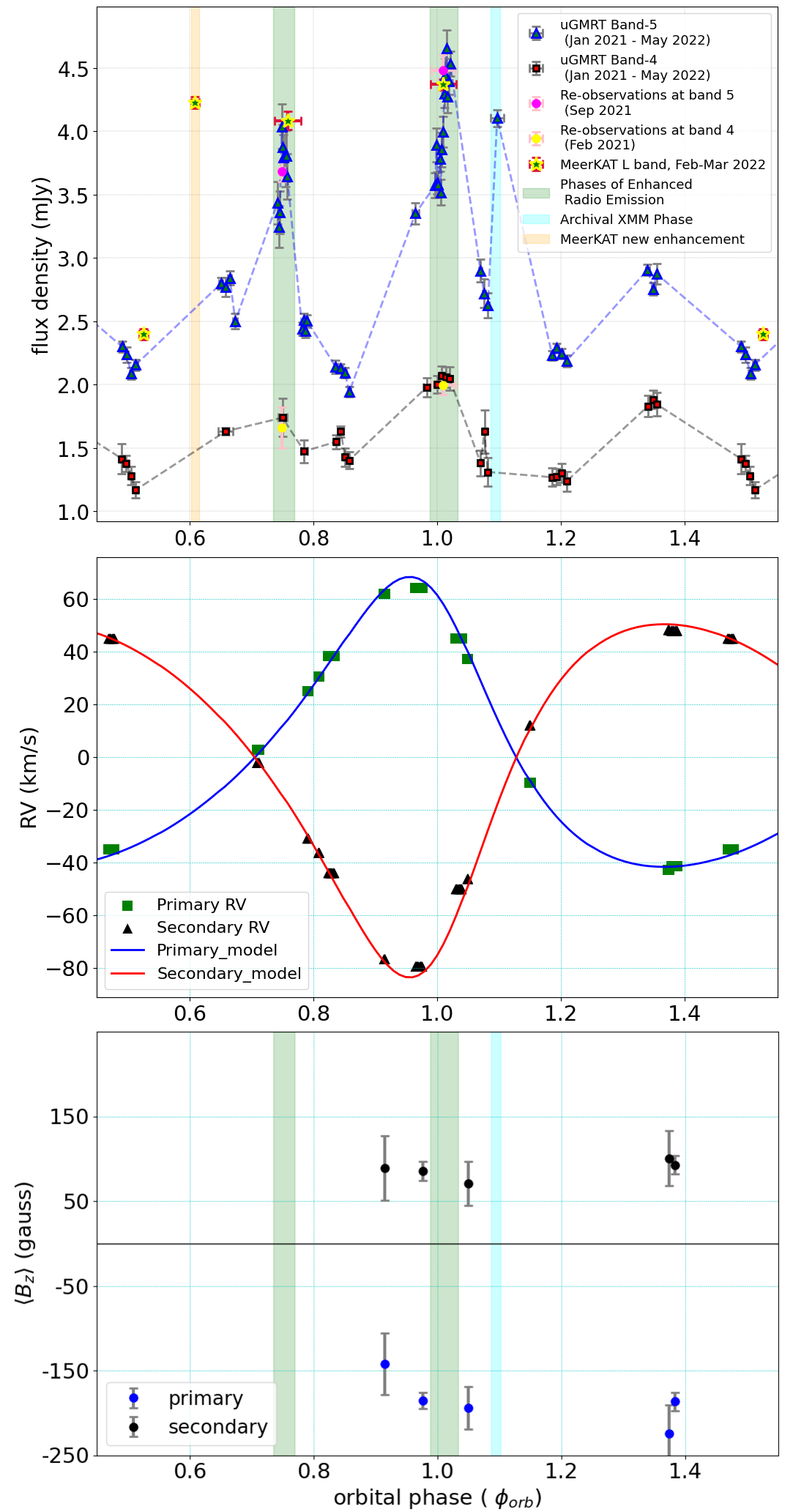}
\caption{\textbf{Top:} { Radio light curve obtained with the uGMRT and  MeerKAT observations. The MeerKAT flux densities are offset corrected. The repeating enhancements are highlighted with green shaded vertical bands. The cyan band represent the phase of the archival XMM-Newton observation and its corresponding radio follow-up; the orange band represent the enhancement at $\sim0.6$ phase observed with MeerKAT. The redundant phase observations confirming the repeating nature of the flux density are shown with pink and yellow points. The yellow stars represent the MeerKAT flux densities in the uGMRT band 5 equivalent frequency range.  The error bars do not include the contribution from error in the absolute flux density scale between different days of observations ($\sim$10\%), since they do not affect the inferences. \textbf{ Middle:} The primary and secondary radial velocities (RVs) of  $\epsilon$ Lupi from the  ESPaDOnS data obtained between 2014 and 2015. Overlaid blue and red curves are the best-fitting orbital model \citep{Pablo2019}. \textbf{Bottom:} The $\langle B_z \rangle$ curve of the system showing non-detection of a magnetic null and lack of variation. The constant positive and negative $\langle B_z \rangle$ for the respective components reflects the anti-alignment of the magnetic axes \citep{Shultz2015}. Reproduced with data from  \citet{Shultz2018Cat}
\label{fig:lc}}}
\end{figure}

We carried out the data reduction  using the XMM–Newton Science Analysis System (SAS, Version: 19.1.0). The tasks `{\it emproc}' and `{\it epproc}' were used respectively to process EPIC-MOS and EPIC-PN data. Data were filtered by selecting events only with a pattern below 12 (for EPIC-MOS) or below 4 (for EPIC-PN). Good Time Intervals (GTI) were calculated with selection criteria: PN rate { $\leq$} 0.4 cts/sec and MOS rate { $\leq$} 0.35 cts/sec. After applying these GTI to our events lists, the remaining exposure times after rejecting bad data are 7.58, 7.60, and 5.60 ks for MOS 1, MOS 2, and PN, respectively. The possibility of pile-up was assessed using the task ‘{\it epatplot}’ but significant pile-up was not predicted.

\section{Results} \label{sec:Results}


$\epsilon$ Lupi was  detected  in both uGMRT bands 4 and 5 at all observed epochs.  The  target was also detected with MeerKAT. The apsidal-motion corrected radio light curve together with radial velocity and $B_{\rm z}$ measurements are shown in Fig. \ref{fig:lc}. The light curve shows the presence of variable radio emission over the full phase of observation. Extreme variability is observed at band 5, characterized by the presence of enhancements of width less than $\sim 0.1$ orbital cycles.

The uGMRT observations covered several orbital phases (from 10 different orbits), whereas the MeerKAT observations were obtained at the periastron phase as well as 3 other orbital phases.  In the MeerKAT observations, the source was detected in Stokes I flux density at all observed phases. However,  the source was undetected in the Stokes Q, U and V images at phases other than the periastron. Considering the 3$\sigma$ upper limits in the respective images, the upper limits of polarization fraction for this phase are 2.07\%, 1.99\%, and 2.15\% respectively for Stokes Q, U and V. During the periastron phase, the search for polarization was done by imaging each scan separately. The source remained undetected in Stokes Q and V images, with $3\sigma$ upper limits of polarization fraction of 0.61\% and 0.49\%.  However, the source was faintly detected in Stokes U (Fig. \ref{fig:meerkat_image}) in the band 1060--1450 MHz, with flux density 118 $\pm$ 30 $\mu$Jy,  giving a polarization fraction of 2.05\%, significant at 3$\sigma$. If we consider the whole MeerKAT L-band (856--1717 MHz), the detection is at $>9\sigma$ significance. For incoherent non-thermal processes, intrinsic linear polarization reflects highly relativistic electrons and generally weak magnetic fields \citep{Dulk1985}. It is possible that the measured fraction of linear polarization is significantly lower than the actual intrinsic polarization produced by the source. This situation may arise either due to the large beam size that may blend small components with higher polarization fractions but different field orientations, or by differential Faraday rotation and Faraday depolarization in circumstellar ionized plasma.

\begin{figure}
    \centering
        \subcaptionbox{\label{fig:speca}}{\includegraphics[width=0.9\linewidth]{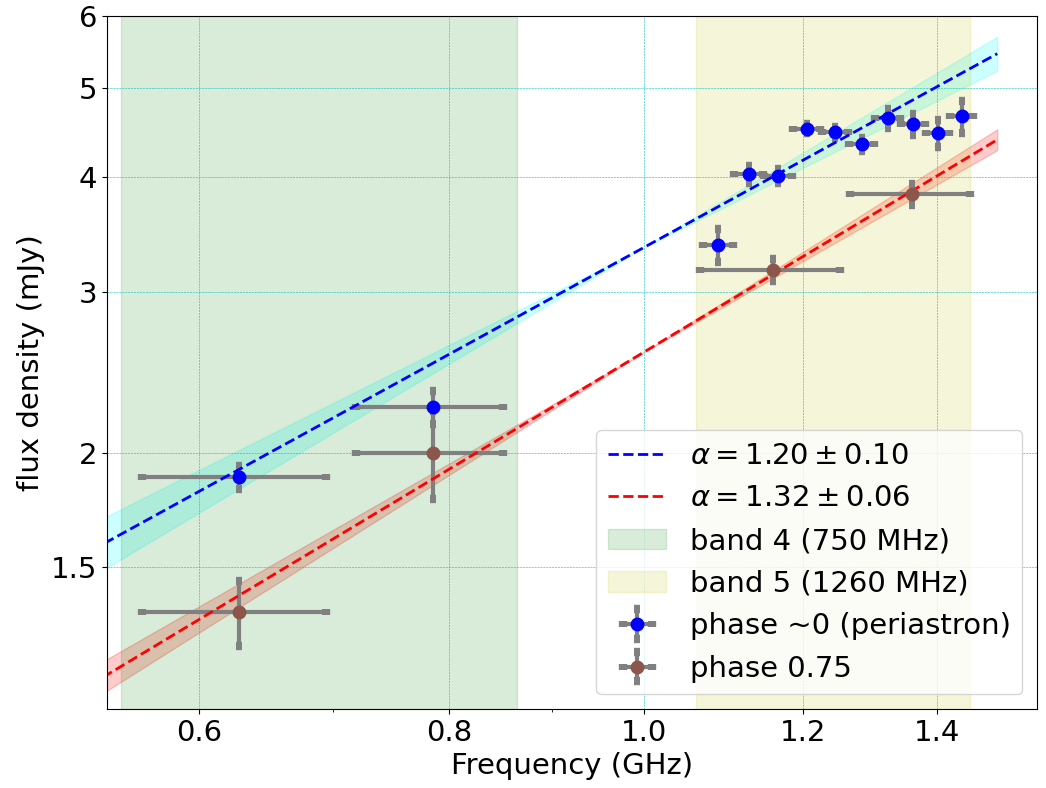}}\par 
        \subcaptionbox{\label{fig:specb}}{\includegraphics[width=0.88\linewidth]{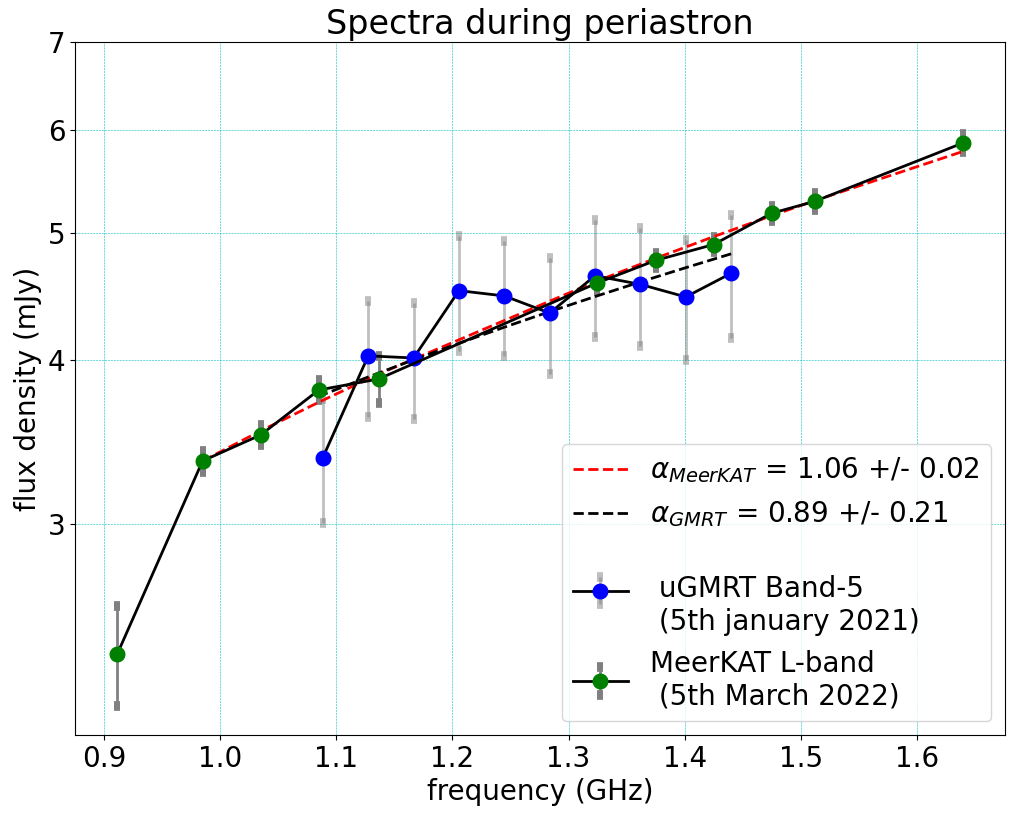}}\par 
\caption{(a) Radio spectra of the source near the enhancement phases from uGMRT observations. { The blue and red line indicate a single power-law fit to the band 5 and band 4 data points, for periastron and phase 0.75, respectively.} (b) The band 5 spectra of $\epsilon$ Lupi from both uGMRT and MeerKAT during the periastron phase suggest a possible cut-off near 1 GHz. \label{fig:alpha}}
\end{figure}

\begin{figure}
\includegraphics[{width=0.48\textwidth}]{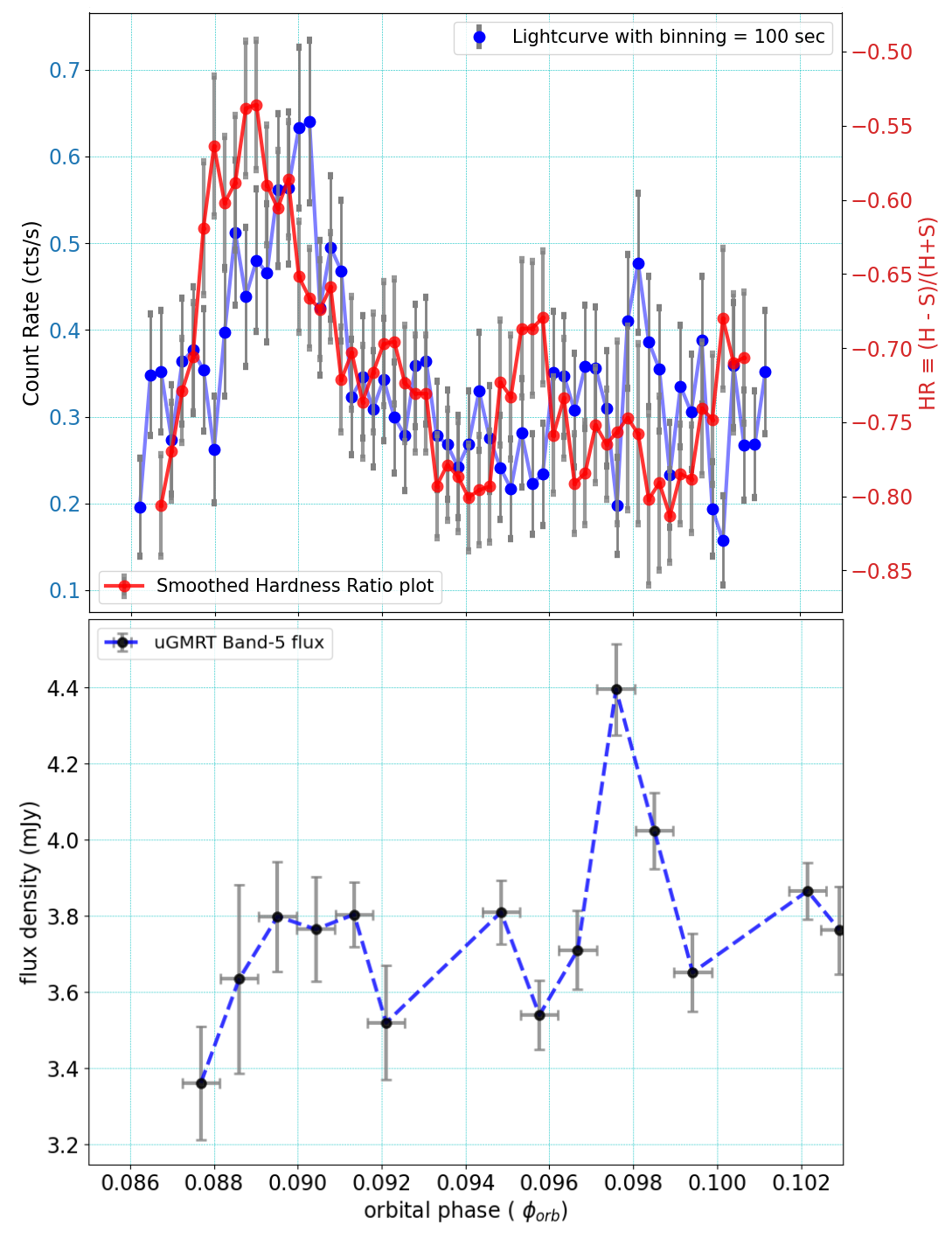}
\caption{\textbf{Top:} The 0.2-10 keV XMM-Newton EPIC-PN X-ray light curve of 
$\epsilon$-Lupi showing variability on short timescales ($\sim$1 hour). {  (blue:) 100-ks binned light curve. (Red:)  The hardness-ratio (HR = (H-S)/(H+S)), where H is 2--10 keV and S is 0.2--2 keV energy ranges. The HR plot mimics the total-counts light curve. Here phase refers to the orbital phase. 
\textbf{Bottom:} Radio light curve obtained from the uGMRT band 5 data observed on 26 May 2022 covering the same orbital phase. The horizontal error bars represent the phase coverage of each data point. The peaks in the X-ray and radio light curves are within the phase error accumulated at the time of the observations due to the uncertainty in apsidal motion ($\Delta \phi_{\rm orb} <0.01$).} \label{fig:xmm}}
\end{figure}

As mentioned above, radio data show two kinds of enhancements, short enhancements or pulses
and smoothly variable possibly periodic emission. Below we explain both in details.

\subsection{Pulses in the light curve} \label{subsec:Results_GMRT}

In uGMRT  band 5 data we observe  three significant enhancements (sharp jumps in the light curve): one near the periastron phase, one at phase 0.75, and another at phase 0.09 (Fig. \ref{fig:lc}). In addition, MeerKAT L band data shows enhancement at phase 0.61. No  sharp enhancements are seen in the uGMRT band 4. The band 5 enhancements at periastron and at phase 0.75 are confirmed to be persistent through our subsequent re-observations of the star at the same orbital phases using both the uGMRT as well as the MeerKAT. The absence of enhancements in band 4 at the periastron phase was also confirmed by re-observing the star at a later epoch. The timescale for the sharp enhancement observed from $\epsilon$ Lupi during the periastron phase is clearly much smaller than the orbital timescale. Subsequent uGMRT observations (taken on 20 July 2021) acquired at phase 0.076, which is just $\sim$8 hrs away from periastron, did not show any elevated flux density level in either band 4 or 5, reinforcing the short duration of these pulses.

To understand the nature of the enhanced radio emission, we investigated the spectral properties of the enhancements at phase 0.75 and at periastron. The periastron phase data with the  uGMRT band 5  and MeerKAT L-band were imaged in several smaller sub-bands to determine the nature of the emission. Between band 4 and band 5 (uGMRT data), the spectral indices at periastron and 0.75 phase are  $\alpha=1.20 \pm 0.10$ and $1.32 \pm 0.06$ respectively, consistent with each other (Fig. \ref{fig:speca}).   The spectral index  of the MeerKAT data during periastron ($\alpha=1.06 \pm 0.02$) is roughly consistent with that of the uGMRT in the frequency range 1.1 to 1.4 GHz. The MeerKAT data also  show a possible cut-off near 1 GHz (Fig. \ref{fig:specb}).

We examined the archival XMM-Newton data for any possible enhancement. The background-corrected light curve from the EPIC-PN instrument is shown in Fig. \ref{fig:xmm}, where we used a 100 sec time-bin. The light curve is further smoothed using the 5-point moving average method. We also measured the hardness ratio between the 0.5-2 keV and 2-10 keV bands for the corresponding phases, which mimics the light curve of the total counts (Fig. \ref{fig:xmm}). We discovered an enhancement in the X-ray light curve at phase $\sim 0.090$. This orbital phase was followed up with uGMRT and a clear enhancement was also observed in the radio light curve in band 5 (Fig. \ref{fig:lc}, top). {  In a very recent study, \cite{Das2023} have reported enhancement in X-ray flux during periastron, as compared to an off-periastron phase. }

\begin{figure*}
    \centering
    \begin{multicols}{2}
        \subcaptionbox{\label{fig:chi1}}{\includegraphics[width=0.87\linewidth]{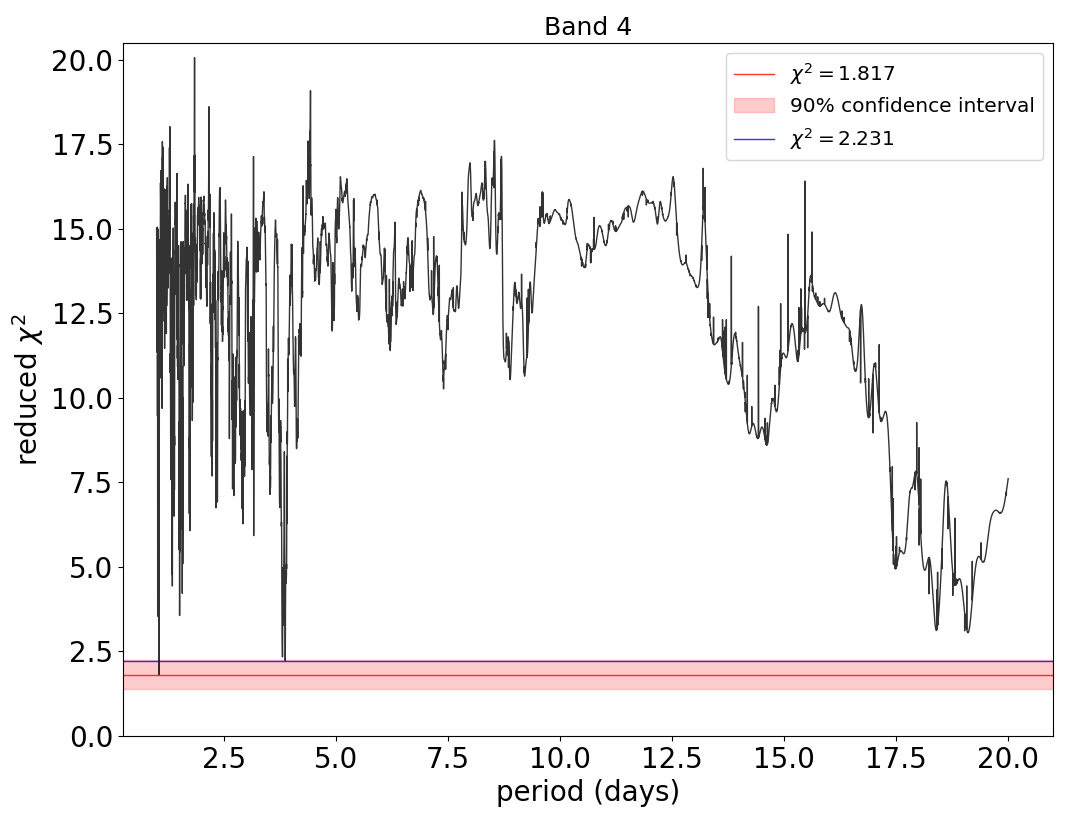}}\par 
        \subcaptionbox{\label{fig:chi2}}{\includegraphics[width=0.87\linewidth]{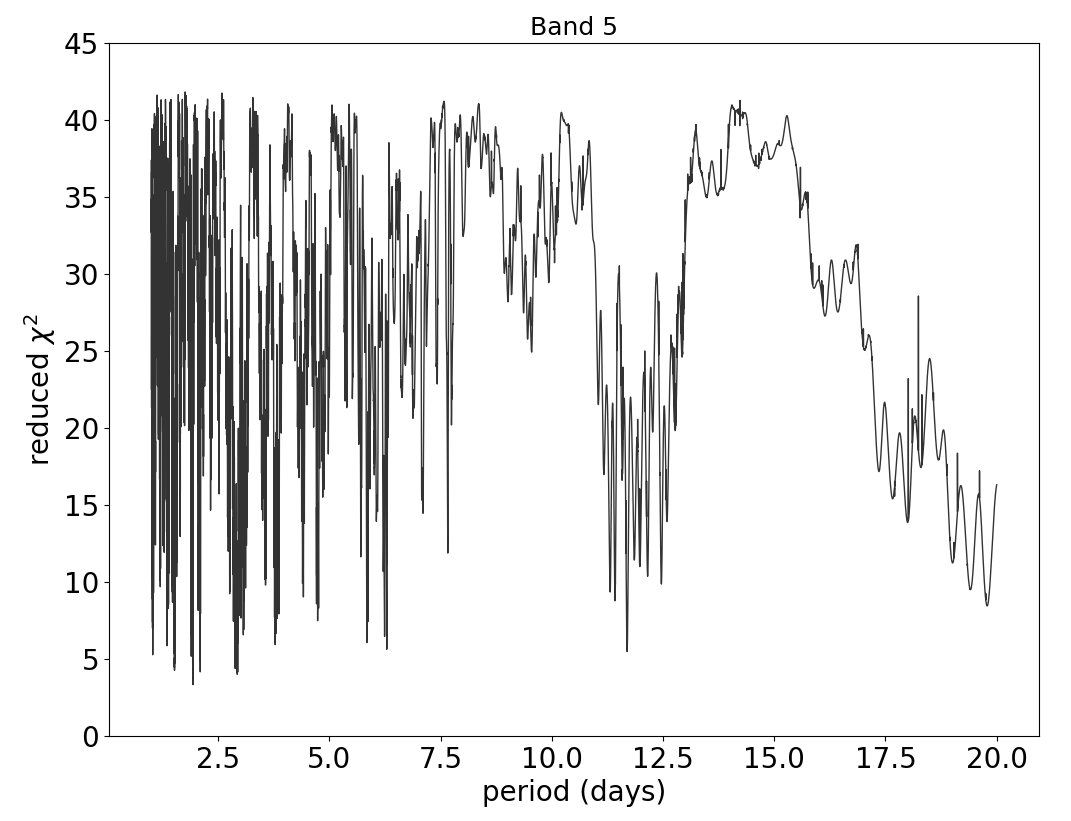}}\par 
    \end{multicols}
    \vspace{-4mm}
\caption{Reduced $\chi^2$ distribution obtained during systematic period analysis for 50000 periods from 1-20 days for band 4 (left) and band 5 (right) data for 
$\epsilon$ Lupi, respectively. The red solid horizontal line in the band 4 plot represent the $\chi^2 = 1.817$ line. The red band represent the 90\% confidence interval for this peak. { As no significant peak was observed from band 5 data, such confidence intervals are not shown in the right image.} \label{fig:chi}}
\end{figure*}

\begin{figure}
\centering
        \subcaptionbox{\label{fig:per4}}{\includegraphics[width=0.92\linewidth]{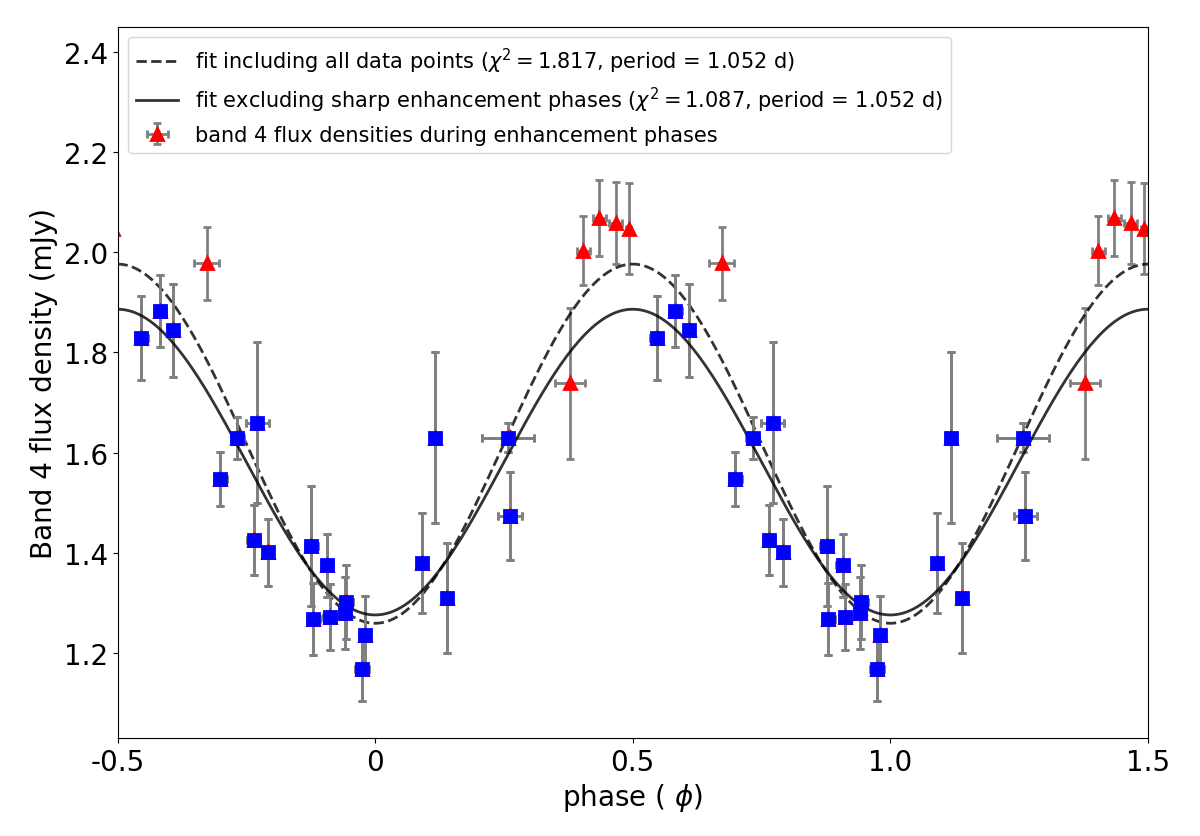}}\par 
        \subcaptionbox{\label{fig:per5}}{\includegraphics[width=0.92\linewidth]{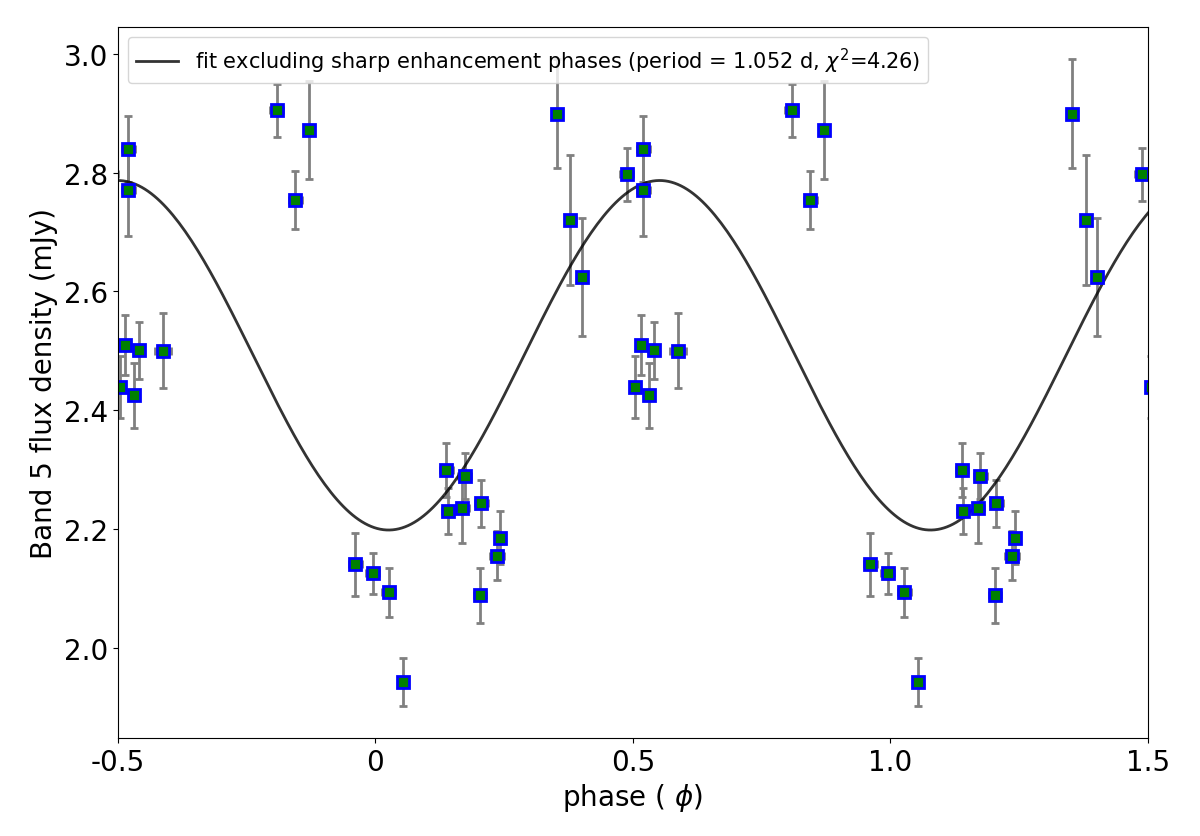}}\par 
\caption{Phase-folded light curve of $\epsilon$ Lupi from GMRT data with period 1.052 days corresponding to the lowest $\chi^2$, (a) for band 4, with full data (red+blue) and data points excluding the enhancement phases that were observed from the band 5 observations (only blue); and (b) for band 5 data points excluding the sharp enhancement phases. \label{fig:band_4_period}}
\end{figure}

\subsection{Periodic variability}

In addition to short enhancements at GHz band, there is a hint of periodic variability in both bands 4 and 5 uGMRT data. 
As the rotation periods of both components are unknown, we carried out a robust analysis of the radio flux density measurements to evaluate periodic variability. Period analysis was performed for 50000 periods from 1 to 20 days, evaluating the reduced $\chi^2$ of a sinusoidal fit to the flux density measurements phased with each period (Fig. \ref{fig:chi}). The best fit from the band 4 data yields a period of 1.052 days with a reduced $\chi^2$ of 1.8 (Fig. \ref{fig:per4}). Another period of 3.87 days obtained from band 4 data has slightly worse $\chi^2$ of $\sim$2.23. As in band 4 no enhancements were found, and also, thermal emission is certain to be absorbed as evident from the free-free radius calculation (see \S \ref{app:ffa}), the emission observed here is expected to be purely gyrosynchrotron coming from the individual stars. Thus the period of 1.054 days may represent the rotation period of one or both of the components (in case the components are synchronized). The band 5 data, on the other hand, yield no similar period. During the period analysis with band 5, the enhancement phases were removed as they resulted in no solution with a reduced $\chi^2$ better than 7. Even after removing the enhancement phases, the band 5 data show poor fit due to fewer data points (Fig. \ref{fig:per5}). Since  the 1.052 day period that seems to represent the band 4 data well does not provide a satisfactory phasing of the band 5 data, the origin of this period is  questionable.

\begin{figure}
    \centering
    \includegraphics[width=0.45\textwidth]{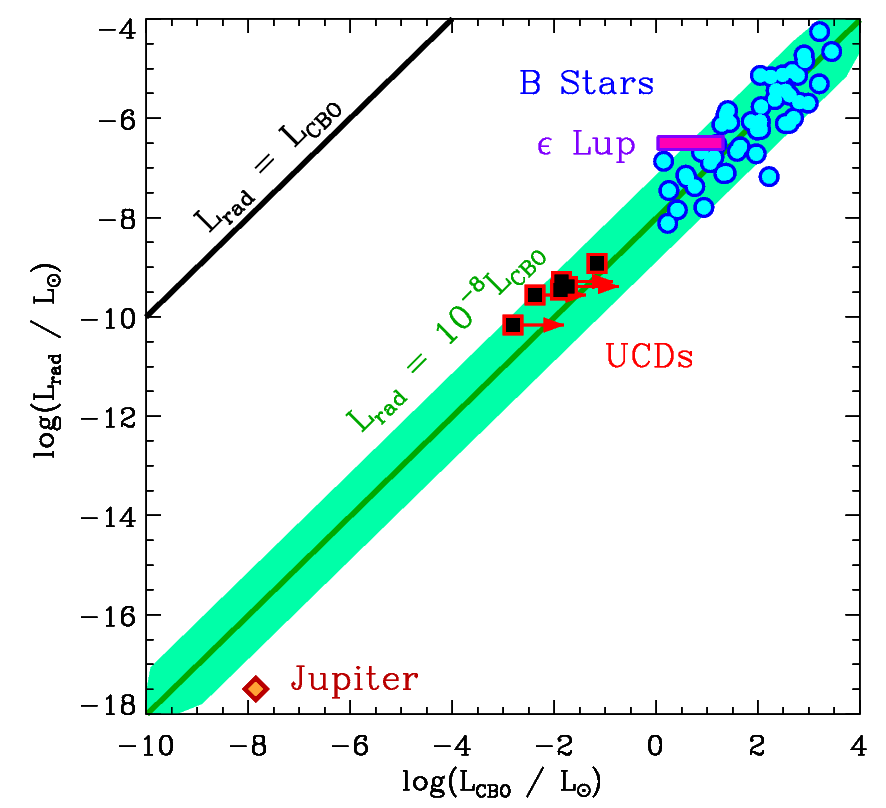}
    \caption[scaling relationship]{Plot of radio luminosity $L_{\rm rad}$ as a function of breakout luminosity $L_{\rm CBO}$ and the position of $\epsilon$ Lupi for different cases. The thick black line indicates $L_{\rm rad}$ = $L_{\rm CBO}$; the solid green line shows the same line shifted so that $L_{\rm rad}$ = $10^{-8} L_{\rm CBO}$. Blue circles indicate B stars, black squares represent ultra cool dwarfs (UCDs), and the yellow diamond shaped point represent Jupiter. The data points are taken from \cite{Leto2021}, and \cite{Shultz2022}. The purple rectangle represent the position of $\epsilon$ Lupi in this scaling relation plot for different conditions: (a) the top of the rectangle represent the $L_{\rm rad}$ of $\epsilon$ Lupi during the periastron phase, while the bottom of the rectangle correspond to the basal $L_{\rm rad}$ of the source. (b) the rightmost point in the rectangle represent the $L_{\rm CBO}$ value corresponding to the period $P_{\rm rot}$=1.052 days, while the leftmost point of the rectangle comes if we assume $P_{\rm rot}=P_{\rm orb} \sim 4.56$ days.}
    \label{fig:my_lcbo}
\end{figure}

Recent studies have found that non-thermal gyrosynchrotron emission from massive stars follows a scaling relation that depends on the magnetic field strength ($B$), rotation period ($P_{\rm rot}$), and radius ($R$) of a given star  \citep{Leto2021, Shultz2022, Owocki2022}. As the radio emission scale with both the magnetic field strength and the stellar rotation rate, we plotted the position of $\epsilon$ Lupi in such scaling relationship plot  for different periods and observed luminosities (Fig. \ref{fig:my_lcbo}). In such a situation, the radio emission is assumed to be gyrosynchrotron, arising according to centrifugal breakout (CBO) model \citep{Owocki2022}, in which the overall total luminosity from CBO events ($L_{\rm CBO}$) in a dipolar stellar field should follow the general scaling relation:

\begin{equation}
    L_{\rm CBO} \approx  \dot{M} \Omega^2 R_*^2 \eta_c^{1/2},
\end{equation}

{ \noindent
where $\eta_c=(B_{\rm d}^2 R_*^2)/(\dot{M} v_{\rm orb})$ is the centrifugal magnetic confinement parameter, $\Omega$ is the rotational frequency, $\dot{M}$ is the mass-loss rate, $B_d$ is the dipolar magnetic field strength of the star, $v_{\rm orb}=\sqrt{G M_* / R_*}$ is the near-surface orbital speed, $R_*$ is the stellar radius, and $M_*$ is the stellar mass}. The radio luminosity ($L_{\rm rad}$) in Fig. \ref{fig:my_lcbo} is defined and calculated in a manner described by \cite{Owocki2022} from band 4 and band 5 data. The purple rectangle in Fig. \ref{fig:my_lcbo} represent the position of $\epsilon$ Lupi in this scaling relation plot for different conditions: (a) the top of the rectangle indicates the $L_{\rm rad}$ of $\epsilon$ Lupi during the periastron phase, while the bottom of the rectangle correspond to the basal $L_{\rm rad}$ of the source. (b) the rightmost point in the rectangle indicates the $L_{\rm CBO}$ value corresponding to the period $P_{\rm rot}$=1.052 days, while the leftmost point of the rectangle comes if we assume $P_{\rm rot}=P_{\rm orb} \sim 4.56$ days.  The basal radio luminosity is exactly consistent with the value expected for the 1.052 days period. Thus radio luminosity certainly cannot rule out the 1.052 days period as being the rotation period of one of the stars. However, as can be seen from the diagram, there is enough scatter in the relation that periods up to about the orbital period are consistent with the scaling relationship. So we can neither confirm, nor rule out this period from the scaling relationship. However, this plot is consistent with the gyrosynchrotron nature from the basal flux.

Assuming rigid rotation and that 1.052\,d is indeed the rotation period of one or both components of the system, the inclination $i_{{\rm rot}}$ of the rotational axis from the line of sight can be determined from the following equation \citep{Stift1976}:

\begin{equation} 
    \sin i = \frac{v \sin i  \cdot  P_{\rm rot}}{50.6 R_*},
\end{equation}

\noindent
where $R_*$ is given in units of solar radii, $v \sin i$ represents the projected rotational velocity in $\rm km s^{-1}$, and $P_{\rm rot}$ is the rotation period in days. We use $v \sin i$ values from \cite{Shultz2018} and {  for radius values, we use the range of values from \cite{Pablo2019} and \cite{Shultz2019} listed in Table \ref{tab:parameters}.  Adopting $P_{\rm rot}= 1.052$ days, and assuming that this period may correspond to either the primary or the secondary, we get the inclination angles to be $8 \pm 4^{\circ}$ for the primary, and $8 \pm 2^{\circ}$ for the secondary, respectively.}  Statistically these low inclination angles are quite unlikely, again making this 1.052 days period questionable. Also, note that the orbital inclination ($i_{\rm orb}$) is about 20$^{\circ}$ \citep{Pablo2019}. From \citealt{Hut1980}, we know that the timescale for spin-orbit alignment is very short in comparison to the circularization and tidal locking timescales; further $\epsilon$ Lupi is a close binary, about middle aged on the main sequence. So we expect $i_{\rm orb} = i_{\rm rot}$. However, a $10^{\circ}$ difference might not be impossible. A denser sampling of the radio light curve may give further insight about the rotational period.

{ 
\section{Discussion} \label{sec:Discussion}


We now examine the nature of radio emission from the $\epsilon$ Lupi system. 
We start with the possibility of radio emission being thermal in nature, produced by the ionized wind.
In this case, we get the mass-loss rate using the relations reported by \citet{Bieging1989} to be $\sim 10^{-6} \ M_{\odot}{\ \rm yr}^{-1}$.  Such a high mass-loss rate is typical of very massive O-type stars and Wolf-Rayet stars, and results in strong emission lines throughout the optical spectrum in both non-magnetic and magnetic examples, which is not observed in this system \citep{Pablo2019, Shultz2019}.
 Furthermore, this mass-loss rate is four orders of magnitude larger than the theoretically predicted value for stars with properties similar to those of $\epsilon$ Lupi ($\sim 0.63 \times 10^{-10} \ M_{\odot}{\ \rm yr}^{-1}$, \citealt{Vink2001, Shultz2019}), implying that the radio emission must be non-thermal. 

Additional evidence of non-thermal radio emission comes from the  brightness temperature $T_{\rm B}$. By considering an emission site with radius as large as the half-path distance during periastron ($\sim$11.3 $R_{\odot}$, \citealt{Pablo2019}), we obtain $T_{\rm B}\approx 1.3 \times 10^{10}\,{\rm K}$. Here, we used the distance d = 156 pc, which was obtained from the Hipparcos parallax \citep{van2007}, as reported by \cite{Pablo2019} and a flux density value of 4.6 mJy (maximum flux density during periastron). Note that this is very likely a lower limit to the actual brightness temperature, since the actual size of the emission site is most probably smaller than assumed. The large value of $T_{\rm B}$ again points to a non-thermal emission mechanism. 

As mentioned in section \ref{subsec:Results_GMRT}, the most striking features of the radio light curve are the sharp enhancements observed at four orbital phases in band 5: nearly coinciding with periastron, at phase $\approx$0.75, at phase $\approx0.61$, and finally at phase $\approx$0.09, corresponding to the X-ray enhancement. 
The enhancements near periastron and phase 0.75 are  persistent, as confirmed  by conducting multiple observations  at different epochs: with the uGMRT as well as with the MeerKAT.  The stability of the sharp enhancements near phases 0.09 and 0.61 cannot be confirmed as multiple observations at this phase have not been conducted. The enhancement observed at periastron is  the strongest enhancement observed. The stability of the spike near
phases 0.09 and 0.61 cannot be confirmed as multiple observations
at this phase have not been conducted.

The characteristics of the observed sharp enhancements in the
radio bands that must be satisfied by the underlying physical scenario/emission mechanism are:
\begin{enumerate}
\item The enhancements are present at four orbital phases mentioned above, one of which is the periastron phase. As the full orbital cycle was not observed, it is possible that more enhancements in the light curve will be revealed in future.
\item The duration of sharp enhancements are $<0.05$ orbital cycles.
\item The lower limit of brightness temperature is found to be $\sim 10^{10}$ K.
\item No circular polarization was detected for the periastron enhancement, but linear polarization of $\sqrt{(Q^2+U^2)} \approx 1.6\%$ (if the full MeerKAT L-band of 856--1717 MHz is considered, or $\approx2.1\%$ if the uGMRT-equivalent band of 1060--1450 MHz is considered) was detected. No polarization was detected at enhancement phases.
\item The sharp enhancements are present only at band 5/L-band, and no enhancements were observed in band 4.
\item  During enhancements, between band 4 and band 5, the spectra have positive spectral indices of $\approx +1$, indicating optically thick emission.
\item For the enhancement near periastron, there is a possible low-frequency cut-off at $\sim 1$ GHz.
\item The enhancement at periastron is stronger than those observed away from periastron.
\item For both persistent enhancements phases, i.e. at periastron and at phase 0.75, multiple observations yield similar flux densities at different epochs.
\end{enumerate}

Characteristics (ii) to (v) give the strongest constraints regarding the radiative processes. The possible underlying physical scenarios can be divided into two cases:

\noindent
\textbf{Case I:} The underlying physical process is triggered only at certain orbital phases, and lasts for a  short time (which determines the duration of the enhancements).

The relevant processes meeting this criteria could be
gyrosynchrotron, synchrotron and/or plasma emission, triggered by either magnetic reconnection or shocks due to wind-wind collision. The high brightness temperature  makes the gyrosynchrotron process unlikely \citep{Robinson1978}. The presence of
stable $\sim 1.6$\% linear polarization across the periastron observation indicates that a quite energetic electron population is present. Such a high brightness temperature, the presence of linear polarization, and the absence of circular polarization points towards synchrotron emission. Furthermore, repetitive behavior
of these enhancements around the periastron phase suggests a role for binarity.  In the following subsection (\S  \ref{subsec:band_5}), we investigate the possibility of synchrotron emission from binary interaction: either due to wind-wind collision or as a result of magnetic reconnection as the underlying physical process for the 
sharp enhancements in band 5/L band.

\noindent
\textbf{Case II:} The underlying physical process is present at all times, but the resulting emission is observable only at certain orbital phases.


For the second type of interaction, there are again two possibilities. The first is that the emission site is not visible to the observer at all orbital phases. However, the fact that the system is viewed nearly `face-on', with the rotational inclination angles of the individual stars likely equal (or possibly even smaller, if the 1.052 day period is rotational) to the small orbital inclination \citep{Shultz2015,Pablo2019} makes this possibility unlikely, because our view of the system geometry doesn’t change significantly during an orbit. The remaining possibility for the second kind of interaction is that the emission is highly directed, in which case the relevant emission mechanism is electron cyclotron maser emission (ECME). 
 In \S \ref{subsec:ecme}, we investigate the possibility of ECME in detail.

\subsection{\textbf{Case I: } Synchrotron origin of the radio enhancements at periastron} \label{subsec:band_5}

\begin{figure}
\includegraphics[{width=0.49\textwidth}]{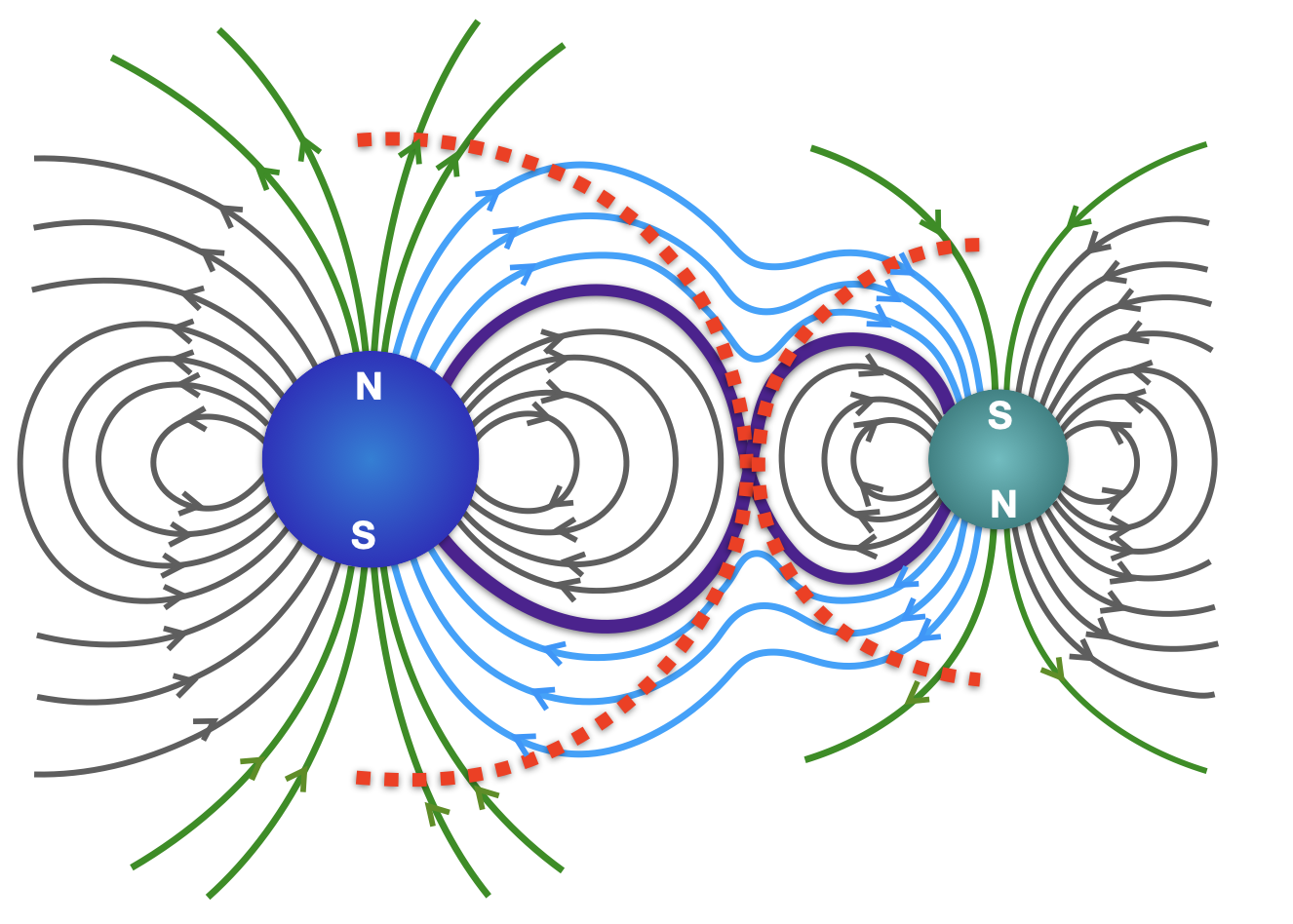}
\caption{ Schematic diagram of magnetic field lines for an $\epsilon$ Lupi-like interacting short period binary system where the components have anti-aligned dipole fields. The dark gray lines represent closed magnetic field lines, and the green lines represent open magnetic field lines. The blue solid lines are the ones that start on one star and close on the other. The purple solid curves mark the boundary between the field lines that close on the same star from which they originate and those that extend to the other star. At their crossing, the magnetic field strength \textbf{B} vanishes. The dotted red semi-circles represent the Alfv{\'{e}}n radii ($R_{\rm A}$).  \label{fig:alfven}}
\end{figure}

Synchrotron emission can arise due to two mechanisms: 1) particles accelerated in shocks created via wind-wind collisions due to proximity of the stars, and/or  2) particles accelerated via magnetic reconnection. 

Wind-wind collision is not very likely in B-star binaries, as due to their low mass-loss rates the winds are not sufficiently strong to produce enhancements in a short period of time. However, in the case of $\epsilon$ Lupi, the proximity during periastron may cause high-density confined winds inside the magnetospheres to collide, resulting in a shorter phase enhancement of synchrotron emission because of the small zone in which winds are interacting. 
The archival X-ray observations can already give some hints in this scenario. Empirically, the ratio between $L_{\rm x}$ and $L_{\rm BOL}$  for a single main sequence non-magnetic B star is about $10^{-7}$ \citep{Berghoefer1997, Becker2007, Naze2014}, where $L_{\rm x}$ and $L_{\rm BOL}$ are the X-ray and bolometric luminosities, respectively. The ratio can be larger in the presence of a magnetic field as X-rays are produced by magnetically confined wind shocks \citep{Babel1997, ud-Doula2014}.
Any strong excess with respect to this ratio could be explained as an interaction between stellar winds in a binary system. 
From the X-ray spectrum, \cite{Naze2014} found for $\epsilon$ Lupi, the value of $\log (L_{\rm x} / L_{\rm BOL})$ to be $-7.16 \pm  0.02$. In this case, the relatively low ratio of the two luminosities close to periastron ($\phi_{\rm orb} < 0.1$) suggests that the X-ray emission is consistent with that seen for solitary magnetic massive stars, and possibly indicates the lack of strong wind-wind interaction. Densely sampled X-ray observations will be critical for evaluating this scenario.

The other possibility is magnetic reconnection. As mentioned earlier, \cite{Shultz2015} predicted the overlapping of the Alfv{\'{e}}n surfaces of the two components throughout the orbital cycle, resulting in strong magnetic reconnection.
The magnetospheres of the components in $\epsilon$ Lupi are guaranteed to collide. 
Due to the eccentric orbit, the magnetospheres of the two stars are further squeezed near periastron, resulting in
magnetic reconnection being  strongest during periastron. This  can produce recurrent enhancements in the light curve. 
In this scenario, the observed emission will arise from the change in energy stored in the composite magnetic field of the system. The enhancements in radio emission during periastron phases of binary systems were previously observed in some proto-star systems, which were attributed to synchrotron emission driven by reconnection due to binary-magnetospheric interaction (e.g. \citealt{Salter2010, Massi2006, Massi2002, Adams2011}). Note that these systems differ from $\epsilon$ Lupi in one important aspect, which is that unlike $\epsilon$ Lupi, the Alfv{\'{e}}n surfaces of the binary components of the proto-star systems do not overlap at all orbital phases. Nevertheless, the oppositely aligned magnetic fields of the two components of $\epsilon$ Lupi certainly favor the magnetic reconnection scenario (e.g. \citealt{Adams2011}). The associated energy release may produce non-thermal electrons which can then emit synchrotron radiation.

However, one of the major challenges in this scenario is explaining the absence of enhancements in band 4. 
Non-thermal radio emission  can be absorbed by three processes: synchrotron self-absorption, free-free absorption, and Razin suppression. The absence of a band 4 enhancement indicates a sharp decrease of emission with frequency, ruling out synchrotron self-absorption. 
While free-free absorption is exponential in nature and band 4 absorption will be higher than that in band 5, it cannot account for the complete absence of sharp enhancements in band 4
(see Appendix \ref{app:ffa}).

Razin suppression \citep{Rybicki1979, Dougherty2003, Erba2022} can lead to the suppression of lower frequency emission, and keep the frequency of maximum brightness temperature constant (see Appendix \ref{app:razin} for details). Razin suppression of the synchrotron emission will scale as $\approx e^{-\nu_{\rm R}/\nu}$, where  $\nu_{\rm R}$ is the Razin frequency which scales as $n_{\rm e}/B$  (Appendix \ref{app:razin}). 
 Taking 1 GHz as the cutoff frequency, we get the number density to be $1.5 \times 10^9$ cm$^{-3}$ (for $B=30$ G, magnetic field along equator $\sim B_0/2 r^3$). Such high densities are possible because the collision of magnetically confined wind plasma implies the compression of that plasma. 
 For the out-of-periastron enhancements, both $n_{\rm e}$ and $B$ will be smaller, which may explain observing a similar 
 cut-off at other orbital phases. Thus Razin suppression is quite a  plausible mechanism to explain the absence of enhancements in band 4.


\subsection{Case II: Electron cyclotron maser emission origin of the radio enhancements at periastron}\label{subsec:ecme}

In this section we consider the second possibility: the coherent Electron Cyclotron Maser Emission (ECME) mechanism (for details, see Appendix \ref{app:ecme}),  produced by mildly relativistic electrons with an unstable energy distribution, provided the local electron gyrofrequency $\omega_{\rm B}$ is larger than the corresponding plasma frequency $\omega_{\rm p}$ \citep[e.g.][]{Treumann2006}. 
For hot magnetic stars, ECME pulses are expected to be visible close to the magnetic null phases \citep[e.g.][]{Trigilio2000}.

In the case of $\epsilon$ Lupi, there are two issues for the ECME origin of radio emission: 1) the orientations of the stellar rotation and magnetic axes are such that the magnetic nulls are never visible for any of the star. 
2) No circular polarization was detected for the enhanced radio emission. 
The lack of circular polarization presents an obvious problem for the ECME scenario, since ECME is generally highly circularly polarized \citep[e.g.][]{Melrose1982}. 
However, the two stars have similar (albeit not identical) surface magnetic fields which are anti-aligned. This configuration could result in the circular polarization of the ECME pulses cancelling out. Nearly zero circular polarization for ECME pulses was indeed observed for some hot magnetic stars \citep[e.g.][]{Das2022c}. However, this requires a large degree of fine-tuning, and it is unlikely that the circular polarization is perfectly cancelled. Rather, there would likely be some residual circular polarization, which might well fall below the sensitivity threshold, especially given the likelihood of absorption within the stellar wind plasma.

The first difficulty may also be resolved if we consider that ECME production in this system is different from that for solitary magnetic stars where the emission is believed to be produced in auroral rings in an azimuthally symmetric manner \citep{Trigilio2011}. In the case of $\epsilon$ Lupi, ECME is likely
triggered by binary interaction, such as magnetic reconnections at the regions where the two magnetospheres overlap. This naturally removes the magnetic azimuthal symmetry of ECME production. This scenario is then more aligned with the active longitude scenario \citep[e.g.][]{Kuznetsov2012} where only a set of magnetic field lines participate in the ECME production, and the emission is directed along the hollow cone surfaces (of half opening angle of $\approx 90^\circ$) centred at these field lines. 

We show in Appendix \ref{app:ecme} that under the scenario that a given magnetic field line is ‘activated’ due to the perturbation from the companion star,
the ECME in bands 5 and 4 are not necessarily observable at the same rotational phase. Besides, even for the cases where ECME at both frequencies are observable at the same rotational phase, the magnetic field lines involved are not necessarily the same. Thus, the non-observation of ECME in band 4 could either be due to not being able to observe the star at the right rotational phase, or that the magnetic field lines that are required to produce ECME at band 4  are not `activated' (the ones that come in contact with the secondary's magnetosphere).
However, we note that since the visibility of the pulses is also dependent on the rotational phase, the ECME scenario will be relevant only if the rotational frequencies are harmonics of the orbital frequency.


\subsection{Presence of Other Enhancements }

So far we have primarily discussed the enhancement at the periastron phase. However, the uGMRT observation revealed a second persistent peak, separated from periastron by $\sim0.25$ cycles or by $\sim 27.3$ hours. Further observation suggested a third peak  near the same phase where the possible X-ray enhancement is observed, and finally a MeerKAT observation suggested a fourth peak at phase 0.61. Although the repeating enhancements during periastron can be explained as an effect of magnetospheric interaction via reconnection, with emission mechanism being either synchrotron or ECME, the origins of the other enhancements are not clear.

As we do not have information about the individual rotation periods, we cannot conclude whether the stars have the same rotation periods. If they are not synchronized, further enhancements can be caused by magnetic reconnection phenomena due to the relative motion of the magnetospheres \citep{Gregory2014}. The possible enhancement seen in the X-ray light curve can also have any of the origins stated above. A dense sampling of orbital phase is thus necessary to fully characterize the variability of the system. Also, due to very slow overall winding of the coupled magnetic ﬁelds of the magnetically coupled system, magnetic energy is gradually accumulated, which may be eventually released in a ﬂare when the instability threshold is reached \citep{Cherkis2021}. Such a scenario is possible in this system. If the rotation period is a harmonic of the orbital period, ECME can be a plausible mechanism to produce such flares.

\section{Conclusion } \label{sec:Conclusion}

In this work we report the discovery of GHz and sub-GHz emission from the close magnetic binary $\epsilon$ Lupi. This is the only known main-sequence binary system to show direct evidence of magnetospheric interaction.  The star is detected in both band 4 (550-950 MHz) and band 5 (1050-1450 MHz) of uGMRT and the  L band  (900 - 1670 MHz) of MeerKAT at all observing epochs.  

The most striking feature of the radio light curve is the existence of sharp radio enhancements at four orbital phases: One during the periastron phase and three others at orbital phases 0.09, 0.61 and 0.75 in bands $>1$ GHz. The enhancement at phase $\sim 0.09$ is also coincident with the enhancement seen in the archival XMM-Newton data. MeerKAT data reveals 1.6--2.1\% Stokes U polarization during periastron. 

We propose that magnetic reconnection is the most favorable scenario for accelerating electrons during periastron. The non-thermal sharply enhanced radio emission can possibly arise via  two mechanisms:
\begin{enumerate}
\item \textit{Synchrotron emission suppressed below 1 GHz via Razin suppression}. In this scenario, especially during periastron, strong magnetic reconnection may take place due to proximity, and recurrent emission may arise from the change in energy stored in the composite magnetic field of the system.
\item \textit{Coherent electron cyclotron maser emission with lower cut-off 1 GHz}. In this scenario, certain magnetic field lines are ‘activated’ due to the perturbation from the companion star, making ECME observable at different rotational phases at different frequencies. However, this scenario required serious  fine tuning. 
\end{enumerate}


The basal emission is consistent with being gyrosynchrotron in nature. We find periodic variability in the basal radio emission with a $\sim1.052$ day period. However, it remains inconclusive whether  this represents  the rotation period of one or both stars in the system. 

While the nature of both the periodic variability of the basal flux and the recurring sharp flux enhancements remain to be determined, these patterns of variation appears to be unique among magnetic hot stars, and are almost certainly related to $\epsilon$ Lupi's status as the only known doubly magnetic hot binary. These tantalizing indications that the system's magnetospheres are in fact overlapping establish $\epsilon$ Lupi as a test system for the physics of magnetospheric binary stars. 

}
\section*{Acknowledgements}

We thank the referee for their detailed and valuable comments. A.B. and P.C. acknowledge support of the Department of Atomic Energy, Government of India, under project no. 12- R\&D-TFR-5.02-0700. B.D. acknowledges support from the Bartol Research Institute. G.A.W. acknowledges support in the form of a Discovery Grant from the Natural Sciences and Engineering Research Council (NSERC) of Canada. M.E.S. acknowledges the financial support provided by the Annie Jump Cannon Fellowship, supported by the University of Delaware and endowed by the Mount Cuba Astronomical Observatory.  The GMRT is run by the National Centre for Radio Astrophysics of the Tata Institute of Fundamental Research. The MeerKAT telescope is operated by the South African Radio Astronomy Observatory, which is a facility of the National Research Foundation, an agency of the Department of Science and Innovation. The National Radio Astronomy Observatory is a facility of the National Science Foundation operated under cooperative agreement by Associated Universities, Inc.

\section*{DATA AVAILABILITY}

The uGMRT data used in this work is available from the GMRT online archive (\url{https://naps.ncra.tifr.res.in/goa/data/search}). The XMM-Newton data are available from the XMM-Newton Science Archive (XSA) (\url{https://www.cosmos.esa.int/web/xmm-newton/xsa}). The MeerKAT data can be found in SARAO MeerKAT archive (see \url{https://skaafrica.atlassian.net/servicedesk/customer/portal/1/article/302546945}). The analyzed data are available upon request to the corresponding author.


\bibliographystyle{mnras}
\bibliography{biswas}



\appendix


\section{Flux-scale Mismatch}

\begin{table*}
\caption{Flux densities of other sources from the MeerKAT periastron observation on 5 March 2022, and uGMRT periastron observation on 5 January 2021. The mean offset ratio is 0.75719.  \label{tab:flux_mismatch}}
\begin{tabular}{ccccc}
\hline
\hline
{RA} &  {DEC} & {uMGRT Flux Den.} &   {MeerKAT Flux Den.} &   {Ratio (uGMRT/MeerKAT)}
 \\ \hline
15:22:36.3 & -44:45:09.4 & 1.163 $\pm$ 0.048 & 1.670 $\pm$ 0.031 & 0.69641 $\pm$ 0.03152 \\
15:23:07.5 & -44:42:32.4 & 0.636 $\pm$ 0.048 & 0.801 $\pm$ 0.041 & 0.79401 $\pm$ 0.07241 \\
15:23:18.8 & -44:44:07.9 & 1.506 $\pm$ 0.061 & 2.385 $\pm$ 0.040 & 0.63145 $\pm$ 0.02768 \\
15:22:38.2 & -44:45:30.7 & 0.630 $\pm$ 0.061 & 0.653 $\pm$ 0.037 & 0.96478 $\pm$ 0.10823 \\
15:22:04.4 & -44:42:36.4 & 0.397 $\pm$ 0.049 & 0.565 $\pm$ 0.025 & 0.70265 $\pm$ 0.09213 \\
15:23:40.5 & -44:37:04.3 & 2.630 $\pm$ 0.110 & 3.684 $\pm$ 0.043 & 0.71390 $\pm$ 0.03100 \\
15:23:43.3 & -44:37:50.3 & 2.462 $\pm$ 0.099 & 3.258 $\pm$ 0.058 & 0.75568 $\pm$ 0.03323 \\
15:23:36.3 & -44:38:44.2 & 1.587 $\pm$ 0.082 & 2.223 $\pm$ 0.038 & 0.71390 $\pm$ 0.03885 \\
15:23:31.2 & -44:40:36.2 & 1.054 $\pm$ 0.066 & 1.958 $\pm$ 0.033 & 0.53830 $\pm$ 0.03491 \\
15:21:58.4 & -44:40:46.1 & 0.818 $\pm$ 0.057 & 0.748 $\pm$ 0.035 & 1.09358 $\pm$ 0.09179 \\
15:23:23.7 & -44:33:18.2 & 0.961 $\pm$ 0.074 & 1.298 $\pm$ 0.040 & 0.74037 $\pm$ 0.06141 \\
15:23:18.9 & -44:47:31.9 & 0.788 $\pm$ 0.060 & 1.063 $\pm$ 0.045 & 0.74130 $\pm$ 0.06458 \\ \hline
\end{tabular}
\end{table*}

We noticed a nearly constant offset between the flux densities of sources in the field of view obtained from MeerKAT and uGMRT. We selected some of the sources from the field of view for which the flux density does not vary significantly within different days obtained from a same telescope. We selected the MeerKAT periastron observation on 5 March 2022, and uGMRT periastron observation on 5 January 2021 to compare the flux densities.  We then obtained the mean offset value (Table \ref{tab:flux_mismatch}) and reduced the MeerKAT flux accordingly. It should be noted that (i) the flux densities of these test sources do not vary significantly among observations with the same telescope, (ii) the flux densities of the uGMRT flux calibrators were in agreement with the VLA calibrator manual{\footnote{\url{https://science.nrao.edu/facilities/vla/observing/callist}}}, and (iii) The MeerKAT L-band flux densities of all sources from our analysis matched well with the results of MeerKAT's SDP pipeline products available in MeerKAT archive.  The reason behind this constant flux-scale offset is not yet known, and we plan to address this issue in future publications.

\section{Razin Suppression} \label{app:razin}

Razin suppression (e.g. \citealt{Rybicki1979, Dougherty2003, Erba2022}) is basically the  suppression of radiation whose refractive index is $<1$ in a given medium. This effect has been used to explain the spectra of solar microwave bursts and their steep slopes at lower frequencies. \citet{belkora96} used the Razin  suppression effect to explain the solar burst spectrum of 16 July 1992 taken with the Owens Valley Solar Array.  They used the X-ray data  of the flare to show that free-free absorption may be present, but is inadequate to explain the shape and evolution of the spectra.  They found that the Razin effect not only leads to the suppression of lower frequency emission, but the frequency of maximum brightness temperature also remains constant.

Here we attempt to explain the effect of Razin suppression and its possible role in the low-frequency cutoff near $\sim$ 1 GHz. The presence of plasma in the emitting region means that the phase velocity of light in the medium is c/$n_{\rm r}$, where $n_{\rm r}$ is the refractive index, which decreases towards low frequencies. As a result, synchrotron radiation is reduced at lower frequencies due to suppression of the beaming effect responsible for synchrotron radiation. The consequence is a low-frequency cutoff near the Razin frequency ($\nu_{\rm R}$) given by \cite{Ginzburg1965}:

\begin{equation}
    \nu_{\rm R} = 2 \times 10^{-8} \frac{n_{\rm e}}{B} \ \ {\rm GHz},
\end{equation}

\noindent
where $n_{\rm e}$ is the electron density in cm$^{-3}$, and $B$ is the magnetic ﬁeld strength in G. The suppression of the synchrotron emission scales as $\approx e^{-\nu_{\rm R}/\nu}$. Taking 1 GHz as the cutoff frequency and 59 G as the magnetic field strength, we get the number density to be $2.95 \times 10^9$ cm$^{-3}$. As the magnetospheres will collide during periastron, the winds confined in the components can reach such high densities. For the out-of-periastron enhancements, both $n_{\rm e}$ and $B$ will be smaller, which may explain observing a similar cut-off at other orbital phases. Thus Razin suppression is quite a  plausible mechanism in determining the structure of the radio enhancements in $\epsilon$ Lupi.

This effect can also explain the positive spectral index observed in this case. In a sufficiently steep distribution of relativistic electrons, Razin suppression of the synchrotron emission becomes more pronounced. This in turn may produce a spectral energy distribution with a power-law slope that mimics the result for thermal emission, particularly in a given waveband \citep{Erba2022}. 

\section{Free-Free Absorption} \label{app:ffa}

To investigate the role of free-free absorption (FFA), we calculated intra-band radio spectra for both bands during peristron phase  and phase 0.75. The band 5 intra-band  spectral index  is  smaller than that in band 4, as well as between band 4 and band 5  (Fig. \ref{fig:alpha}). This is expected if absorption is important. The radius of the free-free radio photosphere ($R_{\rm ff}$, the distance from the star where the free-free optical depth $\tau_{\rm ff}= 1$)  can be calculated using \citet{Torres2011} as:
\begin{equation}
    \tau_{\rm ff} = 5 \times 10^3 \  \dot{M}^2_{-8} V_{\infty '}^{-1}  \nu^{-2} T_{\rm wind}^{-3/2} D_{\rm ff}^{-3},
\end{equation}

\noindent
where, $\dot{M}_{-8}$ is the mass-loss rate in units of $10^{-8} M_{\odot}/{\rm yr}$, $V_{\infty '}$ is the terminal velocity in units of $10^8$ cm/s, $\nu$ is the frequency of observation in GHz, $T_{\rm wind}$ is the wind temperature in units of $10^5$ K, and $D_{\rm ff}$ is the distance from the star, in units of $3 \times 10^{12}$ cm. When $\tau_{\rm ff}= 1$, $D_{\rm ff}=R_{\rm ff}$.
Taking the theoretical mass-loss value ($\sim 0.63 \times 10^{-10} M_{\odot}/{\rm yr}$, \citealt{Shultz2019}), we calculate the free-free radius assuming the parameters $V_{\infty '}$ = 2004 km/s, and $T_{\rm wind} \sim T_{\rm eff}/2 = 11 $ kK. For band 4, we get $R_{\rm ff} \sim 15.7 \ R_{*}$ and for band 5, we get $R_{\rm ff} \sim 11.2 \ R_{*}$, taking a radius of $4.64 R_{\odot}$ \citep{Pablo2019} for the primary. 

Adopting the lower limit value of $R_{\rm A}$ of the primary star to be $\sim 11 \ R_{*}$  \citep{Shultz2019}, we get the ratio $R_{\rm ff}/R_{\rm A} \sim 1.4$ for band 4 and $\sim 1$ for band 5.  $R_{\rm ff}>R_{\rm A}$ signifies that the magnetospheric emission will be hidden within the radio photosphere of the wind, i.e. most of the non-thermal emission will be absorbed. This supports our assumption of heavy free-free absorption in band 4 but not completely in band 5. Although this ratio is higher than most B-type stars \citep{Chandra2015}, due to large uncertainties in the mass-loss and other parameters, this value is not too significant. The fact that more absorption is expected in band 4 remains true irrespective of the uncertainties. While free-free absorption may play an important role, it cannot be solely responsible for the absence of enhancements in band 4.

\section{ECME}\label{app:ecme}

\begin{figure*}
    \centering
    \includegraphics[trim={0cm 5cm 0cm 1cm}, clip, width=0.75\textwidth]{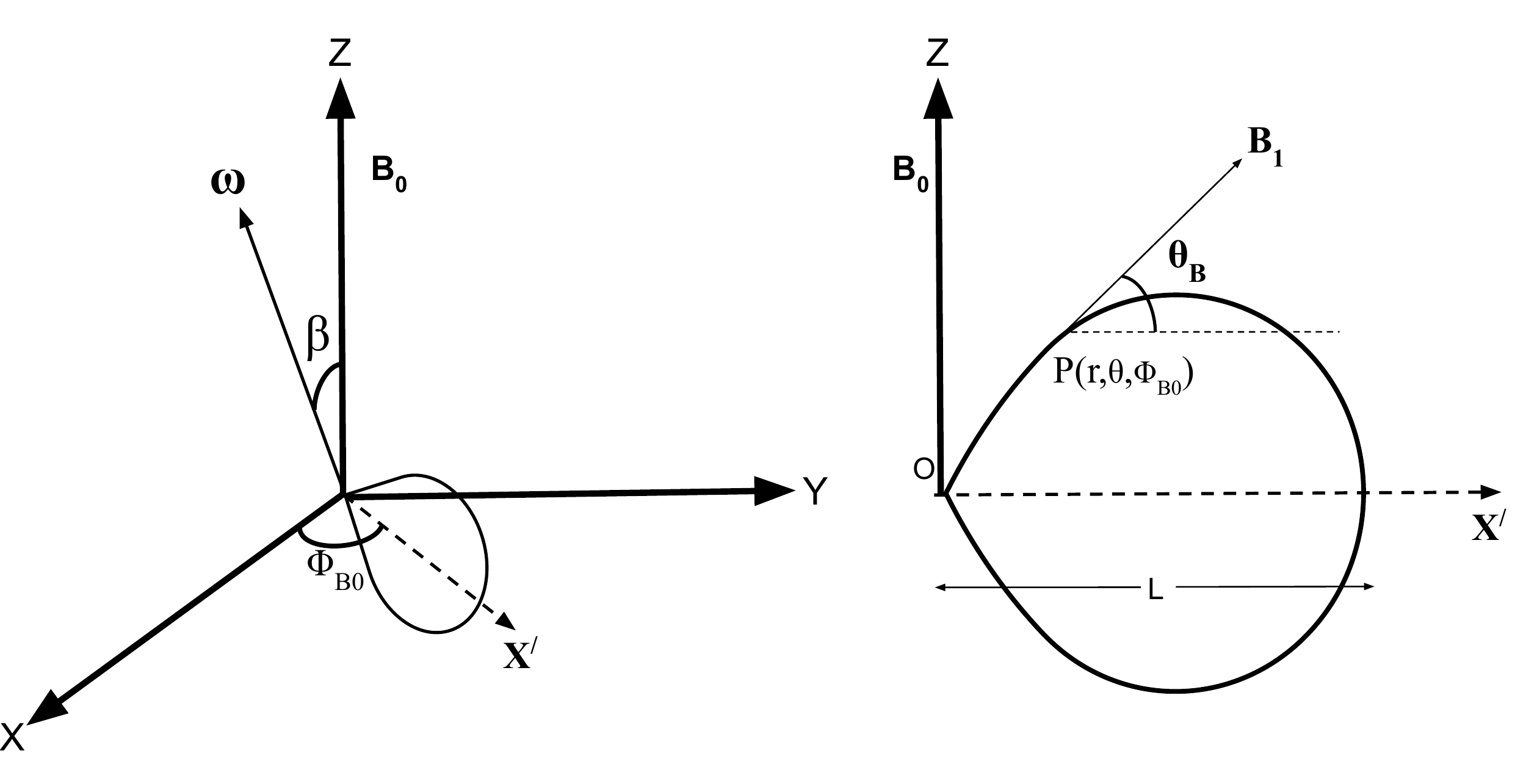}
    \caption{\textit{Left:} The co-ordinate system used for calculating light curves due to ECME produced along certain `active magnetic longitudes' (\S\ref{subsec:ecme}). The plane passing through the dipole axis and the rotation axis is defined as the XZ plane with the Z-axis along the dipole axis. This plane is rotating in the observer's frame of reference. \textit{Right:} The point `P' on the active longitude where the ECME is produced.}
    \label{fig:ecme_active_longitude}
\end{figure*}

ECME gives rise to highly circularly polarized ($\sim 100\%$) emission, directed almost perpendicular to the local magnetic field direction \citep[e.g.][]{Melrose1982}. The frequency of emission is proportional to the local $\omega_{\rm B}$, which results in the fact that in a stellar magnetosphere, higher frequencies are produced closer to the star (where the magnetic field is stronger) and vice-versa.
In this section, we aim to investigate whether there are any circumstances under which ECME emitted at a right angle to the local magnetic field lines can explain the observed radio emission from the $\epsilon$ Lupi system. Considering the similarity between the two components, we perform our calculations for only the primary star. Also, we assume that only one of the magnetic hemispheres (that facing the line-of-sight) is relevant in terms of production of \textit{observable} ECME.

\begin{figure}
    \centering
    \includegraphics[trim={2cm 5cm 2cm 1cm}, clip, width=0.45\textwidth]{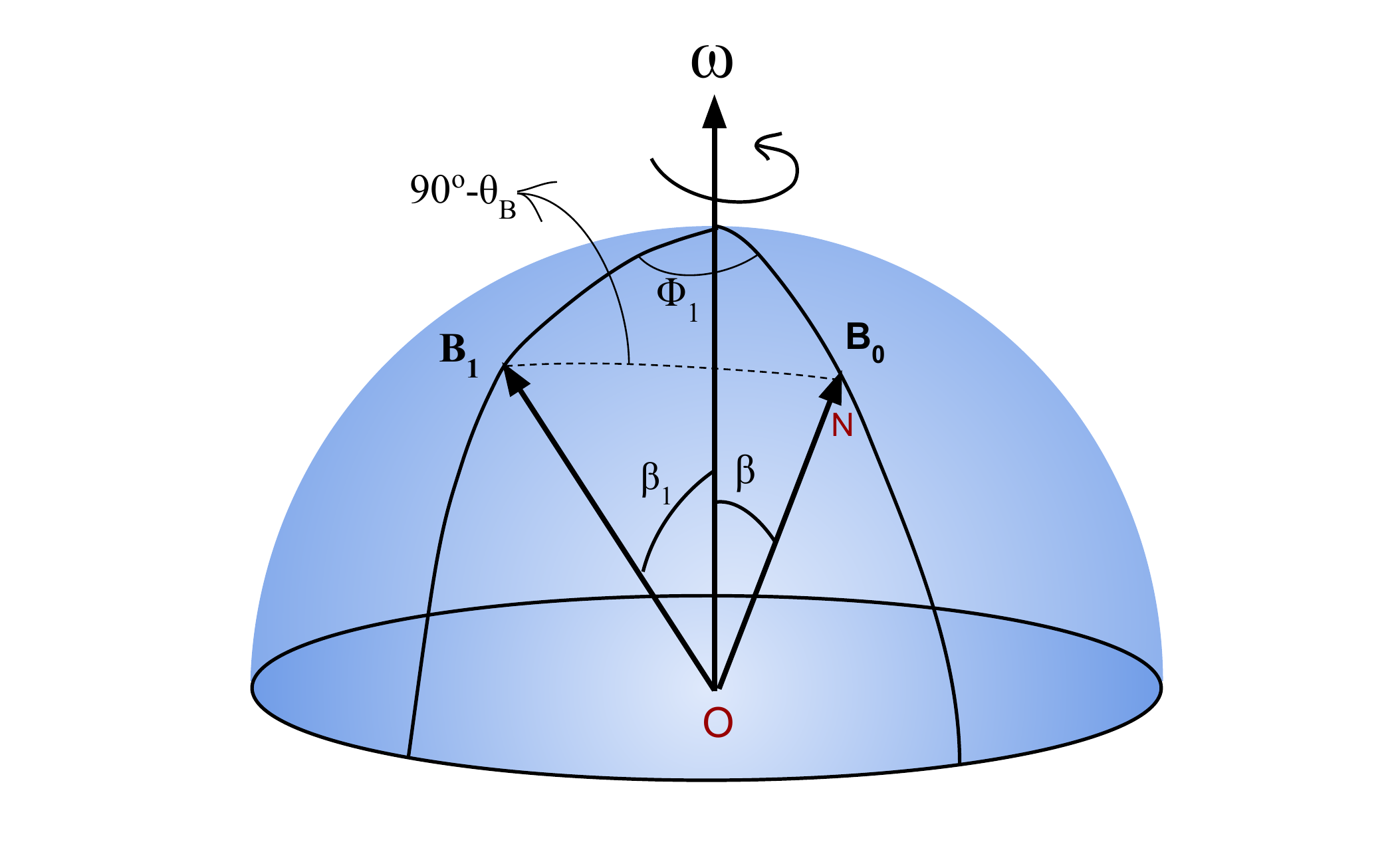}
    \caption{The difference between the rotational longitudes corresponding to the magnetic dipole axis and the magnetic field vector $\mathbf{B_1}$. Here, $\phi_1 = 2\pi\phi_{\rm rot1}$ See \ref{subsec:ecme} for details.}
    \label{fig:B_B1_diff}
\end{figure}

We first define a co-ordinate frame for the star under consideration (the primary star). We define the plane passing through its rotation and magnetic axes as the X-Z plane (left of Figure \ref{fig:ecme_active_longitude}). Thus the magnetic field line intersecting the rotation axis has longitude $\phi_{\rm B}=0^\circ$.  Let us assume that ECME is produced along a magnetic field line with longitude $\phi_{\rm B}=\phi_{\rm B0}$, and with maximum radius $L$ (see right of Figure \ref{fig:ecme_active_longitude}), such that the radial and polar ($\theta$) coordinates of the points lying along the field line are related as:

\begin{align}
    {r=L\sin^2\theta} \label{eq:dipole_line}
\end{align}

Let `P' be the point at which ECME at frequency $\nu$ is produced, and emitted at an angle $90^\circ$. Our aim is to find the angle between the direction of emission and the line-of-sight at a given rotational phase $\phi_{\rm rot}$, which, in this case, is equivalent to obtaining the angle between the local magnetic field vector at point `P' and the line-of-sight.

If the emission happens at harmonic $s$ of the local electron gyrofrequency, the magnetic field modulus at point P is $B_1\approx\nu/(2.8s)$, where $\nu$ is in MHz and $B_1$ is in gauss. The $\theta$ co-ordinate of point P can then be obtained using the following equation:

\begin{figure*}
    \centering
    \includegraphics[width=0.465\textwidth]{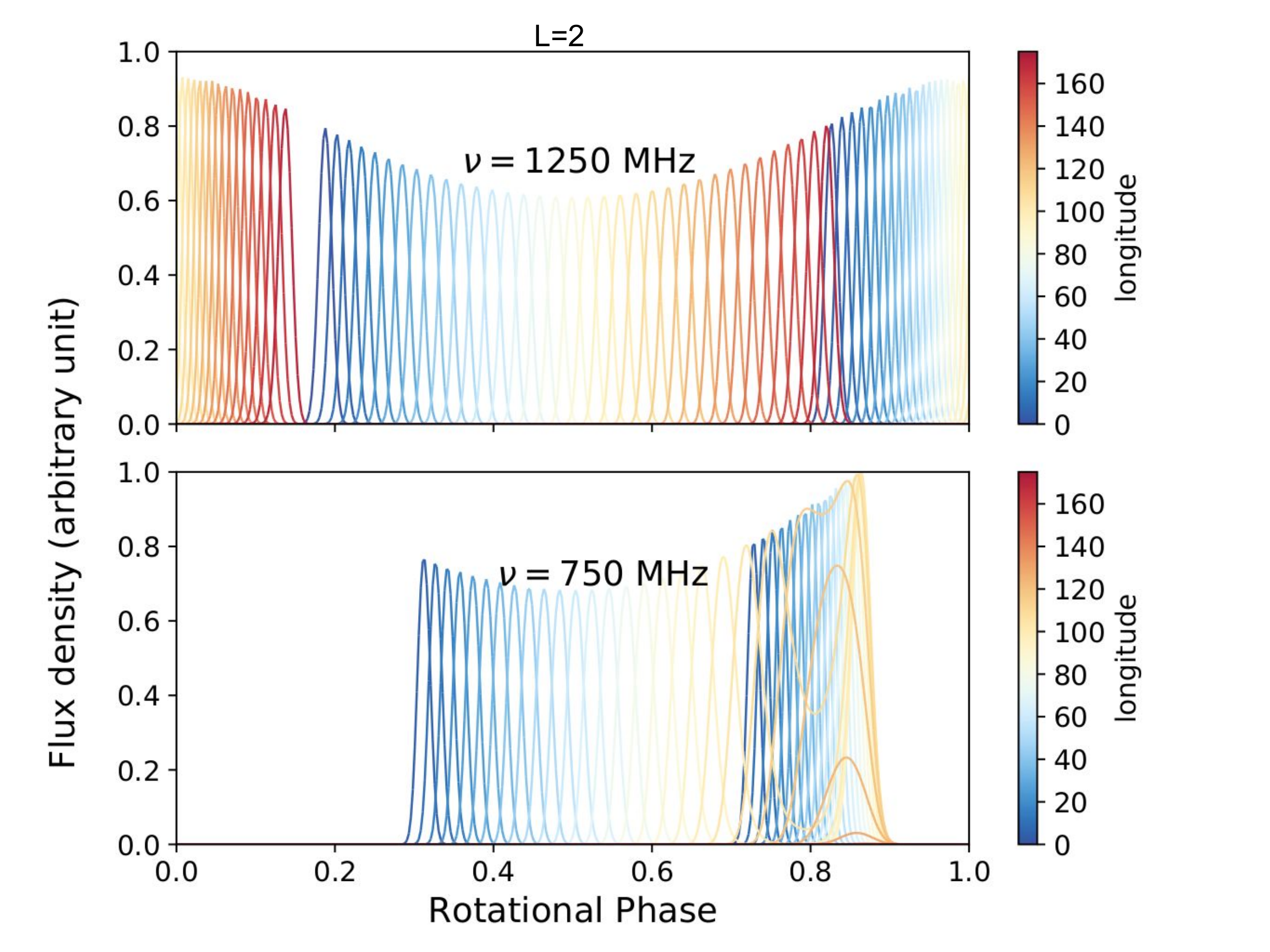}
    \includegraphics[width=0.465\textwidth]{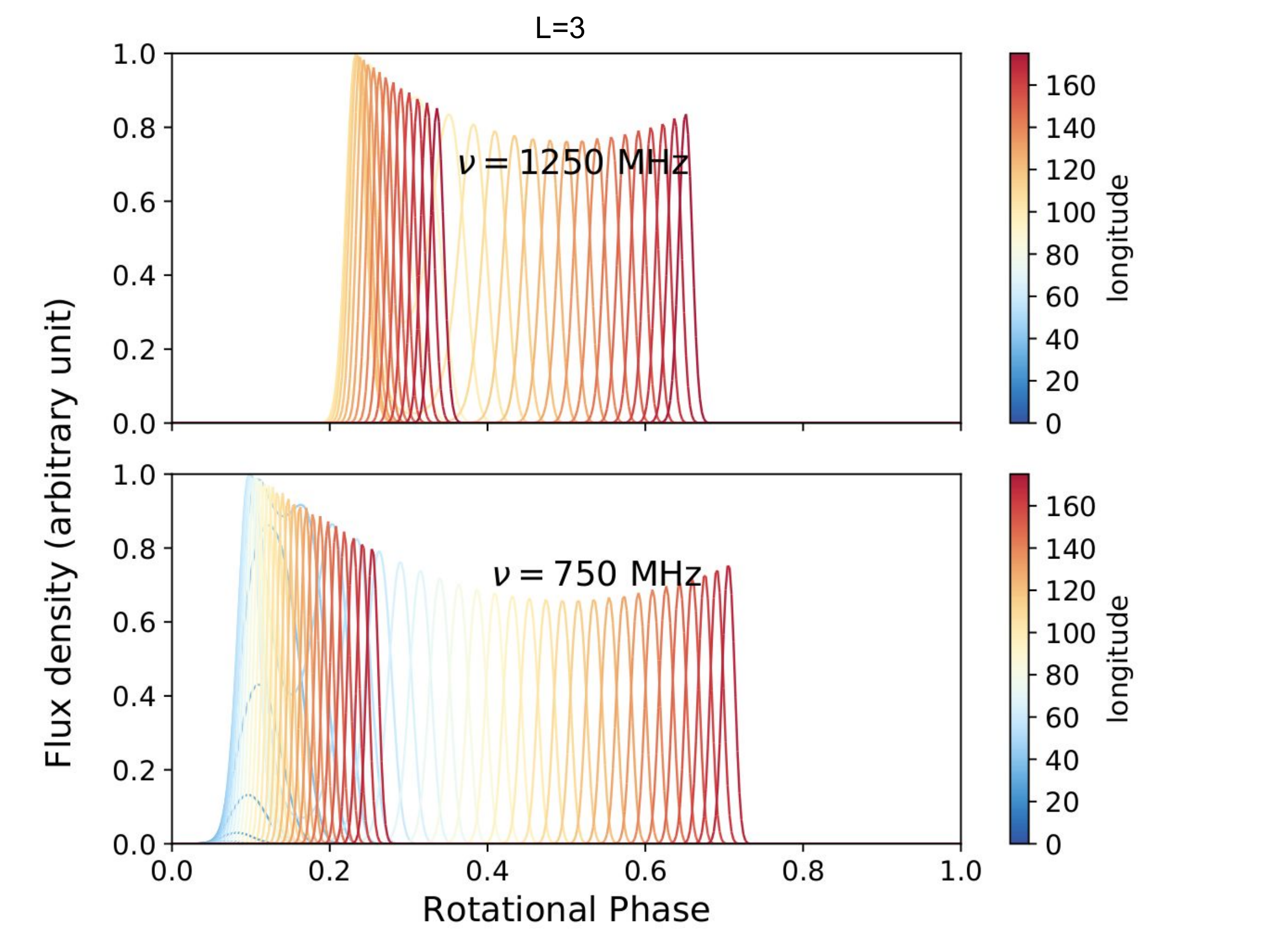}
    \includegraphics[width=0.465\textwidth]{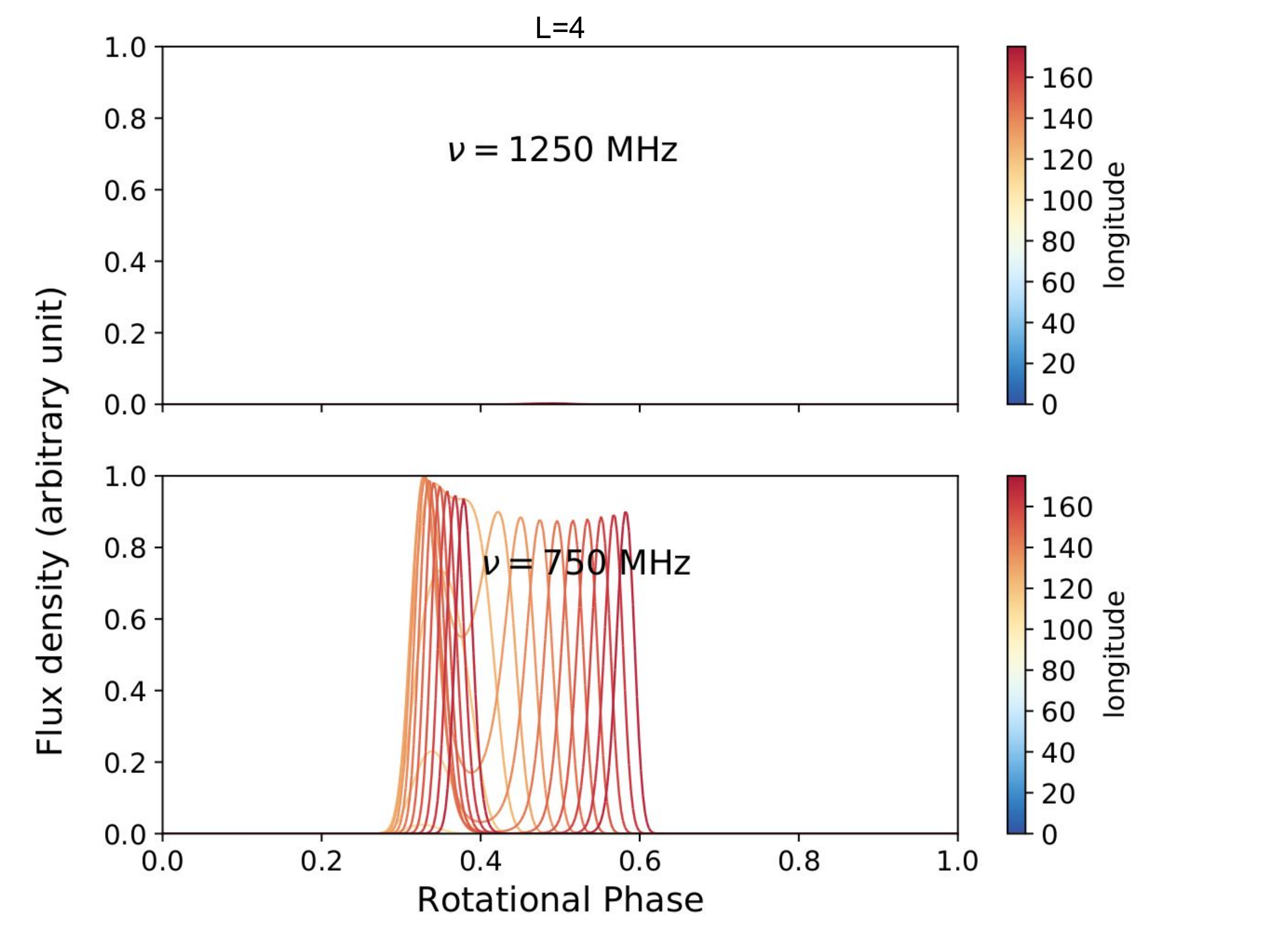}
    \caption{The simulated light curves at 1250 MHz and 750 MHz due to ECME produced at the northern magnetic hemisphere of the primary star of the $\epsilon$ Lupi system, along magnetic field lines with longitudes lying in the range $0^\circ-180^\circ$. We have assumed that the dipole strength is 790 G, and the ECME is emitted at the second harmonic $s=2$ of the local electron gyrofrequency, at right angle to the local magnetic field vector. The three figures correspond to three different values of the quantity $L$ (see Eq. \ref{eq:dipole_line}) in units of the stellar radius.
    Note that the flux densities are in arbitrary units and normalized by the maximum of the combined set of light curves at the two frequencies.}
    \label{fig:ecme_lightcurves_primary}
\end{figure*}

\begin{equation}
    {B_1 = \frac{B_{\rm p}}{2{(L\sin^2\theta)}^3}\sqrt{4-3\sin^2\theta}}  \label{eq:theta_equation}
\end{equation}

\noindent
where $B_{\rm p}$ is the polar field strength. The $r$ co-ordinate can then be obtained using Eq. \ref{eq:dipole_line}. With this information, the vector magnetic field at point P ($\mathbf{B_1}$) can be found, which in turn will allow us to calculate the angle $\beta_1$ between the rotation axis and the magnetic field vector at point P:

\begin{equation}
    {\beta_1=\cos^{-1}(\sin\beta\cos\theta_{\rm B}\cos\phi_{\rm B0}+\cos\beta\sin\theta_{\rm B})}
\end{equation}

\noindent
where $\beta$, called the obliquity, is the angle between the stellar rotational and magnetic dipole axes, and $90^\circ-\theta_{\rm B}$ is the angle between the dipole axis and $\mathbf{B_1}$ (see right of Figure \ref{fig:ecme_active_longitude}). The angle between $\mathbf{B_1}$ and the line-of-sight, at a given rotational phase $\phi_{\rm rot}$ is given by \citep[e.g. Eq. 7 of ][]{Das2020b}:

\begin{align}
    \cos\theta_{\rm B1}&=\cos\beta_1\cos i_{\rm rot}+\sin\beta_1\sin i_{\rm rot}\cos 2\pi(\phi_{\rm rot}+\phi_{\rm rot1}),
\end{align}

\noindent
The zero point of the rotational cycle corresponds to the stellar orientation when the North magnetic pole comes closest to the line-of-sight; $\phi_{\rm rot1}$ is the difference between the rotational longitudes corresponding to the magnetic dipole axis ($\mathbf{B_0}$) and the magnetic field vector $\mathbf{B_1}$, and is given by the following equation (Figure \ref{fig:B_B1_diff}):

\begin{equation}
   \cos 2\pi\phi_{\rm rot1} =\frac{\sin\theta_{\rm B}-\cos\beta\cos\beta_1}{\sin\beta\sin\beta_1}.
\end{equation}

The observer will see the emission whenever $\theta_{\rm B1}\approx 90^\circ$.

Following \citet{Das2020b}, we calculate the flux density $S$ as a function of rotational phase $\phi_{\rm rot}$ using the following equation:

\begin{equation}
    S(\phi_{\rm rot})=\exp\{-{(\theta_{\rm B1}-90^\circ)}^2/2\sigma^2\},
\end{equation}

where $\sigma$ represents the width of the emission beam (taken to be $1^\circ$ here). 

To obtain the ECME light curves from the primary star, we vary the longitude ($\phi_{\rm B0}$) of the `active' magnetic field line between $0^\circ$ and $180^\circ$ ($180^\circ-360^\circ$ is symmetric to this range). We set $B_{\rm p}=790$ G, $i_{\rm rot}=20^\circ$ \citep[][the value of $B_{\rm p}$ is a lower limit]{Shultz2015,Pablo2019}. For the obliquity $\beta$, only an upper limit is available \citep[$\beta<31^\circ$,][]{Shultz2019}. We set $\beta=10^\circ$.

Under the scenario that a given magnetic field line is `activated' due to the perturbation from the companion star, the value of $L$ is likely to be a function of the separation between the two stars. If $L\sim d/2$, where $d$ is the separation between the two stars, $L$ varies between 2 and 4 $R_*$ \citep[using the estimates for the stellar radius, and the major axis and eccentricity of the orbit from][]{Pablo2019}.

Figure \ref{fig:ecme_lightcurves_primary} shows the light curves due to ECME produced at the northern magnetic hemisphere of the primary star for $L=2,3,4\, R_*$. The different colors represent magnetic field lines with different magnetic longitudes. As can be seen, a larger value of $L$ will favor a lower observable frequency of ECME, and vice-versa.

We now consider the observed properties of the radio emission from the system. From Figure \ref{fig:ecme_lightcurves_primary}, it can be seen that the ECME at the two frequencies are not necessarily observable at the same rotational phase. Besides, even for the cases where ECME at both frequencies are observable at the same rotational phase, the magnetic field lines involved are not necessarily the same. Thus, the non-observation of ECME at 750 MHz could either be due to not being able to observe the star at the right rotational phase, or that the magnetic field lines that are required to produce ECME at 750 MHz, are not `activated' (the ones that come in contact with the secondary's magnetosphere).

In the above calculation, the only parameter that is a function of the orbital phase is $L$. We find that the light curves are significantly affected by changes in $L$ (e.g. for $L=4$, ECME at 1250 MHz should not be visible at all), which can explain not being able to observe ECME at all orbital phases. Thus ECME is observable only for certain combinations of $(L, \phi_{\rm rot}$, $\phi_{\rm B0})$. The current data suggest that the enhancements are always observable at fixed orbital phases. Since the visibility of the pulses is also dependent on the rotational phase, the ECME scenario will be relevant only if the rotational frequencies are harmonics of the orbital frequency.

Since we always observed enhancements at 1250 MHz, it suggests that the `activated' magnetic field lines have lower values of $L$. Nevertheless, in the future, it will be important to obtain observations covering the complete orbital cycle of the system so as to investigate if there is any place along the orbit where enhancements at other frequencies are observable.


\bsp	
\label{lastpage}
\end{document}